\documentclass[12pt]{article}
\pdfoutput=1
\usepackage{amsfonts,amsthm,amsmath,amssymb,graphicx,hyperref,color,yfonts}
\usepackage{tikz,cite}
\usepackage{tikz-3dplot}
\newcommand{\bea}{\begin{eqnarray}\displaystyle}
\newcommand{\eea}{\end{eqnarray}}
\def\be{\begin{equation}}
\def\ee{\end{equation}}
\def\beq{\begin{eqnarray}}
\def\eeq{\end{eqnarray}}

\def\half{{1\over 2}}

\begin{document}
\makeatletter
\@addtoreset{equation}{section}
\makeatother
\renewcommand{\theequation}{\thesection.\arabic{equation}}
\vspace{1.8truecm}

{\LARGE{ \centerline{\bf Microscopic Entanglement Wedges}}}  

\vskip.5cm 

\thispagestyle{empty} 
\centerline{{\large\bf Robert de Mello Koch\footnote{{\tt robert@zjhu.edu.cn}} }}

\vspace{.8cm}
\centerline{{\it School of Science, Huzhou University, Huzhou 313000, China,}}

\vspace{.2cm}
\centerline{{\it School of Physics and Mandelstam Institute for Theoretical Physics,}}
\centerline{{\it University of the Witwatersrand, Wits, 2050, }}
\centerline{{\it South Africa }}

\vspace{1truecm}

\thispagestyle{empty}

\centerline{\bf ABSTRACT}

\vskip.2cm 
We study the holographic duality between the free $O(N)$ vector model and higher spin gravity. Conserved spinning primary currents of the conformal field theory (CFT) are dual to spinning gauge fields in the gravity. Reducing to independent components of the conserved CFT currents one finds two components at each spin. After gauge fixing the gravity and then reducing to independent components, one finds two components of the gauge field at each spin. Collective field theory provides a systematic way to map between these two sets of degrees of freedom, providing a complete and explicit identification between the dynamical degrees of freedom of the CFT and the dual gravity. The resulting map exhibits many features expected of holographic duality: it provides a valid bulk reconstruction, it reproduces insights expected from the holography of information and it provides a microscopic derivation of entanglement wedge reconstruction.

\setcounter{page}{0}
\setcounter{tocdepth}{2}
\newpage
\setcounter{footnote}{0}
\linespread{1.1}
\parskip 4pt

{}~
{}~

\section{Introduction}

Collective field theory \cite{Jevicki:1979mb,Jevicki:1980zg} is a promising approach to the AdS/CFT duality \cite{Maldacena:1997re,Gubser:1998bc,Witten:1998qj} as it offers a constructive method to establish the dual holographic gravity theory starting from the given conformal field theory (CFT). The algorithm consists of two steps. The first step is a change of field variables from the original fields to gauge invariant collective fields. This trades the loop expansion parameter ($\hbar$) of the original CFT for\footnote{In this article we focus on vector models. For vector models, the collective fields are bilocal fields obtained by contracting the gauge indices of pairs of fields.} ${1\over N}$. The second step, a change of coordinates, is needed to clarify the gravitational interpretation of the theory. This identifies the CFT coordinates with coordinates of the dual AdS spacetime. The defining fields of the CFT are in irreducible representations of the so(2,$d$) conformal group. The gauge invariant collective fields, given by products of the defining fields, transform in a direct sum of many irreducible representations of so(2,$d$)\footnote{We will work in $d=3$. Irreducible representations of so(2,3) are labelled by the dimension $\Delta$ and spin $s$ of the primary on which the representation is constructed. The free scalar has $\Delta={1\over 2}$ and $s=0$. Denote this representation as $[\Delta,s]=[{1\over 2},0]$. For the vector model example, the collective bilocal fields are products of pairs of the original fields which transform as $[{1\over 2},0]\times [{1\over 2},0]=[1,0]\bigoplus \oplus_{s=1}^\infty [2s+1,2s]$.}. Each irreducible representation corresponds to a bulk field on the AdS spacetime, so the gravity interpretation is simplest in a basis where the so(2,$d$) generators are block diagonal. The change of coordinates of the second step achieves this transformation from the tensor product basis in which the CFT is defined, to the block diagonal basis. The bilocal holography construction, proposed in \cite{Das:2003vw} and further developed in \cite{deMelloKoch:2010wdf,Jevicki:2011ss,Jevicki:2011aa,deMelloKoch:2012vc,deMelloKoch:2014mos,deMelloKoch:2014vnt,deMelloKoch:2018ivk} provides a detailed illustration of this procedure, when the CFT is a free vector model, which is dual \cite{Klebanov:2002ja,Sezgin:2002rt} to higher spin gravity \cite{Vasiliev:1990en,Vasiliev:2003ev}.

Our focus is on the free $O(N)$ vector model in 2+1 dimensions. The single trace primary operators are a scalar with $\Delta=1$ and spinning conserved currents with dimension $\Delta=2s+1$ and spin $2s$ for every positive integer $s$. Not all components of this current are independent. After imposing that the current is symmetric, traceless and conserved there are only 2 independent components of the CFT current at each $s>0$. The single trace primaries, as usual, determine the spectrum of fields of the dual gravity theory. They are dual to a bulk scalar and a set of higher spin gauge fields, one at each spin $2s$ for every positive integer $s$. After fixing the gauge, solving the associated constraint and taking account of the fact that the gauge fields are symmetric traceless fields, we find there are 2 independent components of the gravity gauge fields at each $s>0$. The bilocal holography map is the statement that these degrees of freedom are identical.

The objective of this paper is to discuss bilocal holography with special attention to how it manages to give a complete and detailed mapping of CFT degrees of freedom to gravitational degrees of freedom, to demonstrate that bilocal holography provides a valid bulk reconstruction, to demonstrate how bilocal holography provides an explicit and detailed example of entanglement wedge reconstruction \cite{Czech:2012bh,Headrick:2014cta,Wall:2012uf,Jafferis:2015del,Dong:2016eik,Cotler:2017erl} and how it realizes the principle of the holography of information \cite{Laddha:2020kvp,Chowdhury:2020hse,Raju:2020smc,Raju:2021lwh,SuvratYouTube}, extending and completing the analysis of\footnote{While the light front map employing equal $x^+$ bilocals was derived in \cite{deMelloKoch:2010wdf}, the map employing equal $t$ bilocals is a new result.} \cite{deMelloKoch:2021cni,deMelloKoch:2022sul}. This constitutes compelling evidence in favour of the collective field theory approach to AdS/CFT. 

There are two natural bilocals that can be employed in the construction of collective field theory: equal time and unequal time bilocals. In Section \ref{Bilocals} we discuss the difference between these two. The equal time bilocal is a description in terms of independent components of the higher spin currents, while the unequal time bilocal includes all components of the current. This is a new result and it shows that by formulating the CFT in terms of equal time bilocal fields, one is performing the reduction to independent degrees of freedom. We then review bilocal holography on a light front in Section \ref{lightfront}. The reduction to physical and independent degrees of freedom in the higher spin gravity has been worked out in complete detail in \cite{Metsaev:1999ui}. The reduction to independent degrees of freedom in the CFT is performed by using equal $x^+$ bilocals. The mapping of bilocal holography gives a complete bijection between the independent operators in the CFT and the independent and physical degrees of freedom of higher spin gravity. The final result of Section \ref{lightfront} is to explain how the holography of information and entanglement wedge reconstruction are reflected in the map. In Section \ref{covariant}, we develop an equal time version of bilocal holography, by employing a description of higher spin gravity that respects the Poincar\'{e} subgroup of the boundary CFT, developed in a fascinating paper \cite{Metsaev:2008fs}. In the CFT we employ equal time bilocals. The form of the holographic mapping is completely parallel to the mapping obtained on the light front. We argue that the equal time holographic mapping again provides a complete bulk reconstruction, it codes information into the higher dimensional spacetime exactly as predicted by the holography of information and it can be used to derive the expected entanglement wedge reconstruction. We present a discussion of these results, and our conclusions in Section \ref{conclusions}. This includes a discussion of some open directions as well as a description of how one might approach the holography of matrix models, again within the collective field theory framework.

Finally, we note that similar constructions have recently been developed in \cite{Aharony:2020omh,Aharony:2021ovo,Aharony:2022feg}. These papers employ the first step in the collective field theory algorithm. They use unequal time bilocals and so do not reduce to independent degrees of freedom in CFT. Further, they construct a gravitational interpretation of the theory by using known results, from the harmonic analysis of conformal symmetry, to extract the irreducible representations contained in the bilocal field. Of course, collective field theory does not require a reduction to independent degrees of freedom. For an approach to the unequal time bilocals, demonstrating how the Witten diagram rules are recovered from collective field theory, see \cite{deMelloKoch:2018ivk}. The free field theory, which we discuss here, corresponds to the (unstable) UV fixed point. For discussions treating the (stable) IR fixed point see \cite{Mulokwe:2018czu,Johnson:2022cbe}.

\section{Equal time versus unequal time bilocals}\label{Bilocals}

The defining fields\footnote{We are assuming a real field. The extension needed to consider a complex field is a simple exercise.} of the vector model $\phi^a$ transform in the vector representation of $O(N)$. The $O(N)$ invariant variables are products of pairs of fields with $O(N)$ indices contracted. Each field is at a distinct point so we naturally obtain bilocal fields. Using equal time quantization and a Hamiltonian approach, the dynamics is written in terms of bilocal fields with the fields in the bilocal at distinct spatial locations, but at the same time. Path integral quantization uses bilocals with fields at distinct times and positions. The goal of this section is to discuss the interpretation of equal time versus unequal time bilocal collective fields. 

A useful result for this discussion is the operator product expansion (OPE) which can be used to express the bilocal as a sum over the single trace primary operators of the CFT. For the free $O(N)$ vector model in $d=3$ dimensions, the single trace primary operators include a scalar $j_{(0)}(x)$ of dimension $\Delta=1$ and spinning currents $j_{(2s)}^{\mu_1\cdots\mu_{2s}}(x)$ of spin $2s$ and dimension $2s+1$ for any positive integer $s$. The relevant operator product expansion is \cite{deMelloKoch:2022sul} 
\bea
\sum_{a=1}^N:\phi^a(x+y)\phi^a(x-y):&=&
\sum_{s=0}^\infty\sum_{d=0}^\infty c_{sd} 
\left(y^\mu {\partial\over\partial x^\mu}\right)^{2d}
y_{\mu_1}\cdots y_{\mu_{2s}} j_{(2s)}^{\mu_1\cdots\mu_{2s}}(x)
\label{explicitOPE}
\eea 
where 
\bea
c_{0d}={1\over 2^{2d} (d!)^2} \qquad {\rm and}\qquad
c_{sd}={(2 s)! (4 s-1)!!\over d! 2^{2 d+4 s-1} (d+2 s)!}\qquad\qquad s>0
\eea
and the spinning currents are
\bea 
j_{2s}(y,x)&=&y^{\mu_1}\cdots y^{\mu_{2s}} j_{\mu_1\cdots\mu_{2s}}(x)\cr\cr
&=&\pi \sum_{a=1}^N\sum_{k=0}^{2s}(-1)^k
{:\left(y^\mu {\partial\over\partial x^\mu}\right)^{2s-k}\phi^a (x)
\left(y^\nu {\partial\over\partial x^\nu}\right)^k\phi^a (x):\over
k!(2s-k)!\Gamma (k+{1\over 2})\Gamma(2s-k+{1\over 2})}
\eea
Applying this OPE to the equal time bilocal $\sigma(t,\vec{x}_1,\vec{x}_2)=\phi^a(t,\vec{x}_1)\phi^a(t,\vec{x}_2)$ we easily find
\bea
\sigma(t,\vec{x}_1,\vec{x}_2)
&=&\langle \sigma(t,\vec{x}_1,\vec{x}_2)\rangle+
\sum_{s=0}^\infty\sum_{d=0}^\infty c_{sd} 
\left(y^\mu {\partial\over\partial x^\mu}\right)^d\, 
y_{\mu_1}\cdots y_{\mu_{2s}} j_{(2s)}^{\mu_1\cdots\mu_{2s}}(x)
\label{opeconnection}
\eea
where we have introduced the coordinates
\bea
x_1^\mu &=& x^\mu+y^\mu \qquad x_2^\mu \,\,=\,\, x^\mu-y^\mu\cr\cr
\quad\Rightarrow\quad
x^\mu&=&{1\over 2}(x_1^\mu + x_2^\mu)\qquad
y^\mu\,\,=\,\,{1\over 2}(x_1^\mu - x_2^\mu)\label{comrelcrds}
\eea
From the right hand side of (\ref{opeconnection}) we see that the coordinate $y^\mu$ contracts with indices of the currents. For the equal time bilocal we have
\bea
y^0=0\qquad y^1={1\over 2}(x_1^1-x^1_2)\qquad y^2={1\over 2}(x_1^2-x^2_2)
\eea
This makes it clear that the equal time bilocal only packages currents with spatial polarizations. In contrast to this, the unequal time bilocal, which has $y^0\ne 0$, packages all polarizations of the current.

In Section \ref{UTbilocal}, we describe how conformal transformations are realized on unequal time bilocal fields. In particular, we explain how to recover the familiar generators of the Poinare transformations and dilatations on the conserved currents, from the transformations of the scalar field. In Section \ref{ETbilocal} we argue that the equal time bilocal describes a reduction of the CFT obtained by eliminating components of the conserved current, with the help of the conservation equation. This argument amounts to understanding how conformal transformations are realized on equal time bilocal fields. We start with a discussion of equal $x^+$ bilocals in a light front quantization and argue that equal $x^+$ bilocals describe a reduction of the CFT obtained by eliminating $+$ polarizations of the conserved current. We then argue that an equal time bilocal in a standard equal time quantization describes a reduction of the CFT obtained by eliminating $0$ polarizations of the conserved current.

A final comment is in order. Reducing to independent components in the CFT entails solving the current conservation equation, as well as the traceless and symmetric conditions. In practise it is the solution of the current conservation equation that is non-trivial. Indeed, the traceless condition is preserved by conformal transformations, so that reducing to the  traceless subspace does not entail modifying the generators. The symmetry of the current is simply the statement that certain current components are equal, which can easily be enforced by contracting current indices with a commuting polarization vector. The $y^\mu$ coordinate of the collective bilocal in (\ref{opeconnection}) plays this role. In contrast to this, solving the current conservation equation entails a choice of which polarization will be eliminated, and then a non-trivial modification of the generators of conformal transformations. For this reason, when performing the reduction we focus only on the solution of the current conservation equation. It is this step of the reduction that is achieved by equal time collective fields.

\subsection{Unequal time bilocals}\label{UTbilocal}

We will write the unequal time bilocal
\bea
\sigma(x_1^\mu,x_2^\mu)&=&\sum_{a=1}^N\phi^a(x_1^\mu)\phi^a(x_2^\mu)
\eea
in terms of the coordinates defined in (\ref{comrelcrds}). It is a simple exercise to determine how the bilocal transforms under a conformal transformation, using the known transformation of the free scalar field, as well as the co-product. Expressing the generators in terms of $x^\mu$ and $y^\mu$ we obtain
\bea
P_{\sigma\mu}&=&{\partial\over\partial x_1^\mu}+{\partial\over\partial x_2^\mu}\,\,=\,\,{\partial\over\partial x^\mu}\cr\cr\cr
J^{\mu\nu}_\sigma&=&x_1^\mu{\partial\over\partial x_{1\nu}}-x_1^\nu{\partial\over\partial x_{1\mu}}+x_2^\mu{\partial\over\partial x_{2\nu}}-x_2^\nu{\partial\over\partial x_{2\mu}}\cr\cr
&=&x^\mu{\partial\over\partial x_\nu}-x^\nu{\partial\over\partial x_\mu}+y^\mu{\partial\over\partial y_\nu}-y^\nu{\partial\over\partial y_\mu}\cr\cr\cr
D_\sigma&=&x_1^\mu {\partial\over\partial x_1^\mu}+x_2^\mu {\partial\over\partial x_2^\mu}+1\,\,=\,\,x^\mu {\partial\over\partial x^\mu}+y^\mu {\partial\over\partial y^\mu}+1\cr\cr\cr
a^\mu K_{\sigma\mu}&=&-{1\over 2}x_1\cdot x_1\, a^\mu\,{\partial\over\partial x_1^\mu}+a\cdot x_1\left(x_1^\rho {\partial\over\partial x_1^\rho}+{1\over 2}\right)\cr\cr
&&\qquad -{1\over 2}x_2\cdot x_2\, a^\mu \, {\partial\over\partial x_2^\mu}+a\cdot x_2\left(x_2^\rho {\partial\over\partial x_2^\rho}+{1\over 2}\right)\cr\cr
&=&-{1\over 2}\big(x\cdot x+y\cdot y\big) a^\mu{\partial\over\partial x^\mu}
+a\cdot x\Big( x^\mu {\partial \over\partial x^\mu}+y^\mu {\partial \over\partial y^\mu}
+1\Big)\cr\cr
&&+\,a^\mu x^\nu\Big(y_\mu{\partial\over\partial y^\nu}-y_\nu{\partial\over\partial y^\mu}\Big)+\, a\cdot y\,y^\mu{\partial\over\partial x^\mu}
\label{forK}
\eea
The subscript $\sigma$ on these generators signifies that they are derived using the co-product of the action of conformal transformations on the free scalar field, i.e. they are the action of conformal transformations on the bilocal $\sigma$. To make contact with generators acting on higher spin currents we need to use the OPE result (\ref{opeconnection}), which we write as
\bea
:\sigma(x_1^\mu,x_2^\mu):&=&\sum_{s=0}^\infty\sum_{d=0}^\infty c_{sd} 
\left(y^\mu {\partial\over\partial x^\mu}\right)^d 
y_{\mu_1}\cdots y_{\mu_{2s}} j_{(2s)}^{\mu_1\cdots\mu_{2s}}(x)\cr\cr
&=&\sum_{s=0}^\infty\sum_{d=0}^\infty c_{sd} 
\left(y^\mu {\partial\over\partial x^\mu}\right)^d j_{(2s)}(x,y)\label{OPEres}
\eea
The action of ${\cal G}\in$ so(2,3) on the bilocal field translates into an action on the primary spinning currents as follows
\bea
{\cal G}_\sigma:\sigma(t,\vec{x}_1,\vec{x}_2):&=&{\cal G}_\sigma\,\sum_{s=0}^\infty\sum_{d=0}^\infty c_{sd} \left(y^\mu {\partial\over\partial x^\mu}\right)^d 
y_{\mu_1}\cdots y_{\mu_{2s}}\, j_{(2s)}^{\mu_1\cdots\mu_{2s}}(x)\cr\cr
&=&\sum_{s=0}^\infty\sum_{d=0}^\infty c_{sd} \left[{\cal G}_\sigma,
\left(y^\mu {\partial\over\partial x^\mu}\right)^d\right] j_{(2s)}(x,y)
+\sum_{s=0}^\infty\sum_{d=0}^\infty c_{sd} 
\left(y^\mu {\partial\over\partial x^\mu}\right)^d {\cal G}_\sigma j_{(2s)}(x,y)\cr\cr
&&\label{complicatedTrans}
\eea
It is simple to check that
\bea
\left[{\cal G}_\sigma,
\left(y^\mu {\partial\over\partial x^\mu}\right)^d\right]=0
\eea
for ${\cal G}\in \{P_\mu,J^{\mu\nu},D\}$. Inspecting the action of these generators on the bilocal field (\ref{forK}) we see that these agree with the usual generators acting on the conserved currents.  To explain this agreement note that translations, scalings and Lorentz transformations of $x_1^\mu$ and $x_2^\mu$ correspond to the same transformations in the new coordinates
\bea
x_1^{\prime\mu}=x_1^\mu+a^\mu\quad x_2^{\prime\mu}=x_2^\mu+a^\mu\qquad
&\Rightarrow&\qquad x^{\prime\mu}=x^\mu+a^\mu\quad y^{\prime\mu}=y^\mu\cr\cr
x_1^{\prime\mu}=\lambda x_1^\mu\quad x_2^{\prime\mu}=\lambda x_2^\mu\qquad
&\Rightarrow&\qquad 
x^{\prime\mu}=\lambda x^\mu\quad y^{\prime\mu}=\lambda y^\mu\cr\cr
x_1^{\prime\mu}=\Lambda^\mu{}_\nu x_1^\nu\quad x_2^{\prime\mu}=\Lambda^\mu{}_\nu x_2^\nu\qquad
&\Rightarrow&\qquad x^{\prime\mu}=\Lambda^\mu{}_\nu x^\nu\quad 
y^{\prime\mu}=\Lambda^\mu{}_\nu y^\nu
\eea
In contrast to the generators described so far, the action of $a^\mu K_\mu$ on the bilocal does not agree with the usual generator acting on the spinning current. This is a consequence of the fact that
\bea
\left[(a^\mu K_\mu)_\sigma\, ,\,
y^\mu {\partial\over\partial x^\mu}\right]=-2a\cdot y y^\nu{\partial\over\partial y^\nu}+y^2 a^\mu {\partial\over\partial y^\mu}-a\cdot y\equiv A^{(1)}\label{FrstCmm}
\eea
\bea
\left[A^{(1)}\, ,\,
y^\mu {\partial\over\partial x^\mu}\right]=-2a\cdot y y^\nu{\partial\over\partial x^\nu}+y^2 a^\mu {\partial\over\partial x^\mu}\equiv A^{(2)}
\eea
\bea
\left[A^{(2)}\, ,\,
y^\mu {\partial\over\partial x^\mu}\right]=0
\eea
The fact that (\ref{FrstCmm}) is non-zero implies that the variation of the right hand side of (\ref{OPEres}) is not just the result of the variation of the primary current: it includes variation of the $y\cdot{\partial\over\partial x}$ factor. Thus, reading off the transformation of the conformal current from this equation is not straight forward. The mismatch is further explained by noting that an infinitesimal special conformal transformation in the original coordinates
\bea
x_1^{\prime\mu}=x_1^\mu+2(x_1\cdot b)x_1^\mu -b^\mu x_1\cdot x_1\qquad
x_2^{\prime\mu}=x_2^\mu+2(x_2\cdot b)x_2^\mu -b^\mu x_2\cdot x_2
\eea
does not correspond to an infinitesimal special conformal transformation in the new coordinates
\bea
x^{\prime\,\mu}&=&x^\mu+2b\cdot x x^\mu-b^\mu x\cdot x+2y\cdot b x^\mu-b^\mu x\cdot y\cr\cr
y^{\prime \mu}&=&y^\mu +2x\cdot b y^\mu+2y\cdot b x^\mu -b^\mu x\cdot y
\eea
so we should not expect a match between the generator of special conformal transformations acting on the conserved current and $K_{\sigma\mu}$ derived above.

At the risk of semantic satiation we finally again note that the OPE given in (\ref{OPEres}) implies that the unequal time bilocal packages all components of the conserved CFT currents.

\subsection{Equal time bilocals}\label{ETbilocal}

Higher spin currents $j_{(s)}^{\mu_1\mu_2\cdots\mu_s}(x^\nu)$, have dimension $\Delta =s+1$, are traceless and conserved
\bea
\partial_\mu j_{(s)}^{\mu\mu_2\cdots\mu_s}(x^\nu)&=&0\qquad\qquad\eta_{\mu\nu}j_{(s)}^{\mu\nu\mu_3\cdots\mu_s}(x^\nu)\,\,=\,\,0
\eea
A useful formalism for describing these currents has been developed by Metsaev \cite{Metsaev:1999ui}. The higher spin current is represented as\footnote{The Fock space here is not coming from the second quantization of a field, but rather it is an ``auxiliary Fock space'' that automates the symmetrization of indices and simplifies the tensor calculus.}
\bea
|j_{(s)}(t,\vec{x},a^\mu)\rangle&=&j_{(s)}^{\mu_1\mu_2\cdots\mu_s}(x^\nu) a_{\mu_1}\cdots a_{\mu_s}|0\rangle
\eea
where $a^\mu$ is a bosonic creation operator with $\bar{a}^\mu$ the corresponding annihilation operator so that
\bea
[\bar{a}^\mu,a^\nu]=\eta^{\mu\nu}\qquad \mu,\nu=0,1,2
\eea
As usual we have $\bar{a}^\mu|0\rangle=0$. The conservation equation and traceless conditions are written as
\bea
\bar{a}^\nu\partial_\nu |j_{(s)}(t,\vec{x},a^\mu)\rangle&=&0\qquad\qquad
\bar{a}^\nu\bar{a}_\nu|j_{(s)}(t,\vec{x},a^\mu)\rangle\,\,=\,\,0\label{conseqn}
\eea
The conservation equation can be used to eliminate a component of the current. Our goal is to construct the conformal transformations in this reduced theory, following the discussion in \cite{Metsaev:1999ui}.

Solving the current conservation equation is straight forward in the auxiliary Fock space description. In an equal $x^+$ quantization it is natural to eliminate $+$ polarizations, producing a description in terms of transverse and $-$ polarizations. In this case the solution to the first of (\ref{conseqn}) is given by ($b$ runs over directions transverse to the lightcone)\footnote{As usual, we assume that $\partial^+$ has no zero modes.}
\bea
|j_{(s)}\rangle ={\rm exp}
\left(-a^+\left[{\bar{a}^+\partial^-+\bar{a}^b\partial^b\over\partial^+}\right]\right)
|i_{(s)}\rangle\equiv {\cal P}|i_{(s)}\rangle
\eea
where we have defined an operator ${\cal P}$ and the state $|i_{(s)}\rangle$ is defined by
\bea
|i_{(s)}\rangle=j^{i_1 i_2\cdots i_s}_{(s)} a_{i_1} a_{i_2} \cdots a_{i_s}|0\rangle
\eea
The indices $i_k$ in this last equation run over $-$ and the directions transverse to the light cone so that no components of the current with a $+$ index are packaged in $|i_{(s)}\rangle$. Operators $O$ acting on the original currents $|j_{(s)}\rangle$ become
\bea
\tilde{O}={\cal P}^{-1}\,O\,{\cal P}\label{reducedop}
\eea
when acting on the reduced current $|i_{(s)}\rangle$. This follows by noting that
\bea
O |j_{(s)}\rangle = {\cal P}{\cal P}^{-1}O{\cal P}|i_{(s)}\rangle = {\cal P}\tilde{O}|i_{(s)}\rangle
\eea
Using this rule we construct operators acting in the reduced theory. As an example, the generators of Lorentz boosts in $d$ dimensions are given by\footnote{Bear in mind that it is $\bar{a}^+$ that has a non-trivial commutator with $a^-$.}
\bea
\tilde{J}^{+-}\,\,=\,\,{\cal P}^{-1}J^{+-}{\cal P}\,\,=\,\,x^+{\partial\over\partial x^+} 
- x^-{\partial\over\partial x^-} - a^-\bar{a}^+\label{jpmreduced}
\eea
\bea
\tilde{J}^{+i}\,\,=\,\,{\cal P}^{-1}J^{+i}{\cal P}\,\,=\,\,
x^+{\partial\over\partial x^i}-x^i{\partial\over\partial x^-}-a^i \bar{a}^+
\eea
\bea
\tilde{J}^{-i}\,\,=\,\, {\cal P}^{-1}J^{-i}{\cal P}\,\,=\,\,
x^-{\partial\over\partial x^i}- x^i{\partial\over\partial x^+} + a^-\bar{a}^i
+a^i{\bar{a}^+\partial^-+\bar{a}^b\partial^b\over\partial^+}\label{lcjmi}
\eea
and
\bea
\tilde{J}^{ij}\,\,=\,\,{\cal P}^{-1}J^{ij}{\cal P}
\,\,=\,\,x^i\partial^j-x^j\partial^i+a^i\bar{a}^j -a^j\bar{a}^i
\eea
where $i,j=1,2,\cdots,d-2$ run over directions transverse to the lightcone. We will take $d=2+1$. In this case there is a single direction transverse to the lightcone so that $J^{ij}$ and $\tilde{J}^{ij}$ vanish.

In an equal time quantization it is natural to eliminate temporal ($0$) polarizations. We can verify that\footnote{The division by $\partial^0$ might not well defined since $\partial^0$ may have zero modes. The treatment of these zero modes depends on the boundary conditions adopted. These details do not play any role in our analysis.}
\bea
|j_{(s)}\rangle ={\rm exp}\left(-a^0\left[{\bar{a}^1\partial^1+\bar{a}^2\partial^2\over\partial^0}\right]\right)|i_{(s)}\rangle \equiv {\cal P}|i_{(s)}\rangle
\eea
with
\bea
|i_{(s)}\rangle=j^{k_1 k_2\cdots k_s}_{(s)} a_{k_1} a_{k_2} \cdots a_{k_s}|0\rangle
\eea
solves the first of (\ref{conseqn}). The indices $k_j$ in this last equation only run over the spatial directions. Using (\ref{reducedop}) we construct the reduced boost generators
\bea
\tilde{J}^{0i}\,\,=\,\,{\cal P}^{-1}J^{0i}{\cal P}\,\,=\,\,
x^0\partial^i-x^i\partial^0 - a^i{\bar{a}^j\partial^j\over\partial^0}
\label{reducedboost}
\eea
\bea
\tilde{J}^{ij}\,\,=\,\,{\cal P}^{-1}J^{ij}{\cal P}
\,\,=\,\,x^i\partial^j-x^j\partial^i+a^i\bar{a}^j -a^j\bar{a}^i
\label{reducedrot}
\eea

The reduced Lorentz generators have a clear interpretation. Lorentz boosts mix temporal and spatial polarizations of the current. In contrast to this, the reduced Lorentz boosts given in (\ref{reducedboost}) mix only the spatial polarizations of the current packaged in $|i_{(s)}\rangle$. To understand what the reduced generators are doing, consider the $s=2$ current and consider an infinitesimal boost along the $i=1$ direction, of rapidity $\epsilon$. The purely spatial component of the current $j_{(2)}^{12}$ (for example) transforms as $j_{(2)}^{12}\to j_{(2)}^{\prime 12}$ with
\bea
j_{(2)}^{\prime 12}&=&j_{(2)}^{12}+\epsilon\left( x^1{\partial\over\partial t}+t{\partial\over\partial x^1}\right)j_{(2)}^{12}+\epsilon j_{(2)}^{02}\label{otrans}
\eea
so that purely spatial components mix with temporal components. This transformation law (as well as the transformation of all other components of the current, and currents with $s\ne 2$) is reproduced by the boost generator
\bea
J^{ab}&=&x^a{\partial\over\partial x_b}-x^b{\partial\over\partial x_a}+M^{ab}
\qquad M^{ab}\,\,=\,\, a^a\bar{a}^b-a^b\bar{a}^a
\eea
in the usual way
\bea
|j'_{(s)}\rangle &=& (1+\epsilon J^{01})|j_{(s)}\rangle
\eea
Using the generator of boosts in the reduced theory $\tilde{J}^{0i}$
\bea
|i_{(s)}^{\prime}\rangle&=& (1+\epsilon \tilde{J}^{01})|i_{(s)}\rangle
\eea
we find
\bea
j_{(2)}^{\prime 12}&=&j_{(2)}^{12}+\epsilon\left( x^1{\partial\over\partial t}+t{\partial\over\partial x^1}\right)j_{(2)}^{12}-\epsilon {\partial_1 j_{(2)}^{12}+\partial_2 j_{(2)}^{22}\over \partial_t}\label{ntrans}
\eea
As we have already noted, this transformation rule does not involve temporal components of the current. Comparing (\ref{otrans}) and (\ref{ntrans}) it is clear that $j_{(2)}^{02}$ in (\ref{otrans}) is replaced by $-(\partial_1 j_{(2)}^{12}+\partial_2 j_{(2)}^{22})/\partial_t$ in (\ref{ntrans}). This replacement rule is implied by the conservation equation
\bea
\partial_t j_{(2)}^{02}+\partial_1 j_{(2)}^{12}+\partial_2 j_{(2)}^{22}=0
\eea
demonstrating that the reduced boost is indeed obtained by eliminating temporal polarizations with the current conservation equation. There is a parallel discussion for the boosts obtained by eliminating $+$ polarizations in an equal $x^+$ quantization.

In the remainder of this subsection we will argue that this reduction is naturally implemented by the equal time bilocal collective field, which is given by
\bea
\sigma(t,\vec{x}_1,\vec{x}_2)&=&\phi^a(t,\vec{x}_1)\phi^a(t,\vec{x}_2)
\eea
To see that this must be the case, note that from the OPE (\ref{opeconnection}) we know that the equal time bilocal packages only spatial polarizations of the current. Since any conformal transformation takes an equal time bilocal into another equal time bilocal, it must be that the collective field performs the reduction outlined above. 

We can check this slick argument with explicit examples. Consider an infinitesimal boost along the $x^1$ direction, generated by $J^{01}$. Perform the Lorentz transformation on the two scalars and then use (\ref{opeconnection}) to work out the transformation implied for the conserved currents. The boost takes $t\to t+\epsilon x^1$, $x^1\to x^1+\epsilon t$ and $x^2\to x^2$. Applying this rule to each scalar field we have\footnote{Recall that the coordinates $x^\mu$ and $y^\mu$ were defined in (\ref{comrelcrds}).}
\bea
(1+\epsilon J^{01})\eta(t,\vec{x}_1,\vec{x}_2)
&=&:\phi^a(t+\epsilon x^1_1,x^1_1+\epsilon t,x^2_1)\phi^a(t, \vec{x}_2):
+:\phi^a(t,\vec{x}_1)\phi^a(t+\epsilon x^1_2, x^1_2+\epsilon t,x^2_2):\cr\cr
&=&\phi^a(t,\vec{x}_1)\phi^a(t,\vec{x}_2)+\epsilon y^1 (:\partial_t\phi^a(t,\vec{x}_1)\phi^a(t,\vec{x}_2):-:\phi^a(t,\vec{x}_1)\partial_t\phi^a(t,\vec{x}_2):)\cr\cr
&&\qquad+\,\,(\epsilon x^1\partial_t+\epsilon t \partial_{x^1}):\phi^a(t,\vec{x}_1)\phi^a(t,\vec{x}_2):
\eea
Now, using the identity (\ref{equaltimeidentity}) derived in Appendix \ref{collectiveidentity}, this becomes
\bea
\epsilon J^{01}\eta(t,\vec{x}_1,\vec{x}_2)=
\epsilon(x^1\partial_t+t \partial_{x^1}+y^1
{\partial_{y^1} \partial_{x^1} + \partial_{y^2}\partial_{x^2}\over \partial_t})
:\phi^a(t,\vec{x}_1)\phi^a(t,\vec{x}_2):
\eea
so that we read off
\bea
J^{01}\,\,=\,\,x^1\partial_t+t \partial_{x^1}
+y^1{\partial_{y^1}\partial_{x^1}+\partial_{y^2}\partial_{x^2}\over \partial_t}
\eea
in perfect agreement with (\ref{reducedboost}). A completely parallel argument shows that $J^{02}$ also comes out correctly. Under the action of $J^{12}$ we have
\bea
\eta(t,\vec{x}_1,\vec{x}_2)\to 
\eta(t,x_1^1-\epsilon x_1^2,x_1^2+\epsilon x_1^1,x_2^1-\epsilon x_2^2,x^2_2+\epsilon x_2^1)
\eea
This transformation is realized using the differential operator
\bea
J^{12}\,\,=\,\, x^1\partial_{x^2}-x^2\partial_{x^1}+y^1\partial_{y^2}-y^2\partial_{y^1}
\eea
in perfect agreement with (\ref{reducedrot}). Together these results reproduce the description of the reduced currents obtained by eliminating $0$ polarizations.

A parallel analysis shows that the equal $x^+$ bilocal reproduces the reduction obtained by eliminating light like polarizations. The equal $x^+$ bilocal is
\bea
\eta(x^+,x_1^-,x_1,x_2^-,x_2)=:\phi^a(x^+,x^-_1,x_1)\phi^a(x^+,x^-_2,x_2):
\eea
The boost generated by $J^{+-}$ generates the transformation\footnote{Here we use of the coordinates defined in (\ref{comrelcrds}) which are given by $x_1=x+y$, $x_2=x-y$, $x_1^-=x^-+y^-$ and $x_2^-=x^--y^-$. The inverse transformation is $x={1\over 2}(x_1+x_2)$, $y={1\over 2}(x_1-x_2)$, $x^-={1\over 2}(x_1^-+x_2^-)$ and $y^-={1\over 2}(x^-_1-x^-_2)$.}
\bea
(1+\epsilon J^{+-})\eta(x^+,x_1^-,x_1,x_2^-,x_2)
&=&\phi^a(x^++\epsilon x^+,x^-_1-\epsilon x_1^-,x_1)\phi^a(x^+,x^-_2,x_2)\cr\cr
&&\qquad +\,\,\phi^a(x^+,x^-_1,x_1)
\phi^a(x^++\epsilon x^+,x^-_2-\epsilon x^-_2,x_2)\cr\cr
&=&\left(1+\epsilon\left(x^+ \partial_{x^+} - x^-\partial_{x^-}-y^-\partial_{y^-}\right)\right) \eta(x^+,x_1^-,x_1,x_2^-,x_2)\cr\cr
\Rightarrow && J^{+-}=x^+{\partial\over\partial x^+}-x^-{\partial\over\partial x^-}-y^-{\partial\over\partial y^-}
\eea
in complete agreement with (\ref{jpmreduced}). The argument for $J^{-i}$ is similar. Finally, under the action of $J^{-i}$ we have

\bea
\epsilon J^{-i}\eta(x^+,x_1^-,x_1,x^-_2,x_2)
&=&:\phi^a(x^+-\epsilon x_1,x_1^-,x_1+\epsilon x_1^-)\phi^a(x^+,x_2^-,x_2):\cr\cr
&&\qquad +\,\, :\phi^a(x^+,x_1^-,x_1)
\phi^a(x^+-\epsilon x_2,x_2^-,x_2+\epsilon x_2^-):\cr\cr
&=&\epsilon y(:\phi^a(x^+,x_1^-,x_1)\partial_+\phi^a(x^+,x_2^-,x_2):-:\partial_+\phi^a(x^+,x_1^-,x_1)\phi^a(x^+,x_2^-,x_2):)\cr\cr
&+&(\epsilon x^-\partial_x-\epsilon x\partial_+
+\epsilon y^-\partial_y):\phi^a(x^+,x_1^-,x_1)\phi^a(x^+,x_2^-,x_2):
\eea
Using the identity (\ref{xplusidentity}) derived in Appendix \ref{collectiveidentity}, we easily find
\bea
J^{-i}=x^-\partial_x- x\partial_{x^+} + y^-\partial_y+y{\partial_{y^-} \partial_{x^+} 
+ \partial_y\partial_x\over \partial_{x^+}}
\eea
in complete agreement with (\ref{lcjmi}).

\section{Bilocal Holography on a Light Front}\label{lightfront}

Gauge theory/gravity duality relates an ordinary quantum field theory to a theory of quantum gravity. It implies that the physical degrees of freedom of these two theories must match. Bilocal holography is an explicit demonstration of this fact and it establishes a precise bijection between the physical and independent degrees of freedom of the two theories. By working in light cone gauge in the higher spin gravity, it is possible to completely gauge fix the theory, to solve the constraint associated to light cone gauge and then to reduce to non-redundant physical degrees of freedom. Denoting the higher spin gauge fields by $A_{(s)}^{\mu_1\cdots\mu_s}$ light cone gauge sets
\bea
A_{(s)}^{+\mu_2\cdots\mu_s}=0\qquad s=2,4,6,\cdots
\eea
As usual, the equations of motion associated to fields set to zero by the gauge condition must be imposed as constraints. The constraints associated to this gauge condition determine all fields with $-$ polarizations $A_{(s)}^{-\mu_2\cdots\mu_s}$. The higher spin gauge fields, which are usually double traceless, become traceless in this gauge. Thus, the physical degrees of freedom are given by a traceless symmetric field $A_{(s)}^{i_1\cdots i_s}$ where the indices take the values $Z,X$. The number of independent symmetric tensors $A_{(s)}^{i_1\cdots i_s}$ is $N^{A_{(s)}}_{\rm symm}=s+1$ so that the number of independent symmetric and traceless tensors is
\bea
N^{A_{(s)}}_{\rm symm,tr}=s+1-(s-2\,+1)=2
\eea
The reduction to physical and independent degrees of freedom in higher spin gravity has been carried out in detail in \cite{Metsaev:1999ui}. We will simply review the results we need in Section \ref{reducehsg}.

The gauge invariant CFT currents $j_{(s)}$ packaged by the bilocal are traceless symmetric and conserved. The number of independent symmetric tensors $j_{(s)}^{\mu_1\cdots \mu_s}$ is $N^{j_{(s)}}_{\rm symm}={1\over 2}(s+1)(s+2)$ so that the number of independent symmetric and traceless tensors is
\bea
N^{j_{(s)}}_{\rm symm,tr}={1\over 2}(s+1)(s+2)-{1\over 2}(s-2\,+1)(s-2\,+2)=2s+1
\eea
The number of symmetric, traceless and conserved tensors is
\bea
N^{j_{(s)}}_{\rm symm,tr,com}=2s+1 -\left(2(s-1)+1\right)=2
\eea
The non-trivial element of this reduction to physical degrees of freedom is the solution of the current conservation equation. This is described in Section \ref{reducecft} and it is accomplished by employing equal $x^+$ bilocal fields.

Bilocal holography establishes the identity of the two physical and independent components of the higher spin gravity gauge field and the two independent components of the spinning CFT current. This mapping is described in Section \ref{lightfrontholography}. In Section \ref{HMapComments} we draw some general lessons from the holographic map.

A comment on notation: we use little letters ($x^+,x^-,x$) for CFT$_3$ coordinates and capital letters ($X^+,X^-,X,Z$) for the coordinates of AdS$_4$.

\subsection{Collective Field Theory}\label{reducecft}

The conformal field theory dynamics is expressed as the collective field theory of an equal $x^+$ bilocal field
\bea
\sigma(x^+,x_1^-,x_1,x_2^-,x_2)=\sum_{a=1}^N\phi^a(x^+,x^-_1,x_1)\phi^a(x^+,x^-_2,x_2)
\eea
The change to collective (invariant) variables ensures that the loop expansion parameter of the resulting field theory is ${1\over N}$ which matches the loop expansion parameter of the dual gravity theory. The reduction to independent components of the current entails solving the conservation equation to eliminate all $+$ polarizations of the current. This reduction is automatically achieved, as explained in Section \ref{ETbilocal}, by using equal $x^+$ bilocal fields. In particular, the generators of conformal transformations are those of the reduced theory.

The change to bilocal field variables necessarily involves a Jacobian, which has been described, for example, in \cite{deMelloKoch:1996mj}. Expanding this Jacobian about the leading large $N$ value of the bilocal generates an infinite sequence of interaction vertices. The Feynman diagram loop expansion using these vertices then reproduces the $1/N$ expansion of correlation functions. These vertices will reproduce the complete non-linear interactions of gravity and they can be compared directly to the completely gauge fixed higher spin gravity theory. In the discussion which follows we work in the large $N$ limit so that these vertices will not play a role.

In formulating the map to the dual gravity theory, it is convenient to Fourier transform from $x^-$ to $p^+$. Thus, we work with the bilocal field
\bea
\sigma(x^+,p_1^+,x_1,p_2^+,x_2)&=&\int \, dx_1^-\,\int \,dx_2^-\,
e^{ip_1^+x_1^-+ip_2^+x_2^-}\sigma(x^+,x_1^-,x_1,x_2^-,x_2)\label{FTbi}
\eea
The equation of motion for this bilocal field is
\bea
i\partial_+\sigma (x^+,p_1^+,x_1,p_2^+,x_2)=\left(-{1\over 2p_1^+}{\partial^2\over\partial x_1^2} -{1\over 2p_2^+}{\partial^2\over\partial x_2^2}\right)\sigma (x^+,p_1^+,x_1,p_2^+,x_2)\label{CFTeom}
\eea

The bilocal field $\sigma$ develops a large $N$ expectation value. Expanding about this leading configuration we have
\bea
\sigma(x^+,p_1^+,x_1,p_2^+,x_2)=\sigma_0(x^+,p_1^+,x_1,p_2^+,x_2)+{1\over\sqrt{N}}\eta(x^+,p_1^+,x_1,p_2^+,x_2)\label{fluctuation}
\eea
It is the fluctuation $\eta(x^+,p_1^+,x_1,p_2^+,x_2)$ that is identified with the fields of the higher spin gravity. Notice that $\eta(x^+,p_1^+,x_1,p_2^+,x_2)$ is a function of 5 coordinates. Finally, the large $N$ expectation value $\sigma_0$ is nothing but the leading large $N$ equal $x^+$ two point function of the field $\phi^a$.

\subsection{Higher Spin Gravity}\label{reducehsg}

The light-cone gauge description of higher spin gravity has been developed in detail by Metsaev \cite{Metsaev:1999ui}. We are interested in the gravity dual to the large $N$ limit of the CFT, which corresponds to free bulk fields. In this case we can work with the Fronsdal description~\cite{Fronsdal:1978rb} rather than the full Vasiliev theory\cite{Vasiliev:1990en,Vasiliev:2003ev}. The spin-$s$ Fronsdal field $A_{\mu_1 \mu_2 \cdots \mu_s}$ is symmetric and obeys a double tracelessness condition
\bea
A{_\nu}{^\nu}{_\rho}{^\rho}{^{\mu_5\cdots \mu_s}} = 0
\eea
The dual to the free $O(N)$ vector model involves these gauge fields, one for every even spin $2s$. The higher spin gauge symmetry is
\bea
A^\prime{^{\mu_1 \cdots \mu_s}} = A^{\mu_1 \cdots \mu_s} + \nabla^{(\mu_1} \Lambda^{\mu_2 \cdots \mu_s)}
\eea
The gauge parameter $\Lambda^{\mu_1 \dots \mu_{s-1}}$ is symmetric and traceless and $\nabla_\mu$ is the AdS covariant derivative. The AdS vierbein $e^A_\mu$ converts frame indices to spacetime indices. In the Poincar\'{e} patch of AdS we have
\bea
e^A_\mu = \frac{1}{z} \delta^A_\mu \, .
\eea
We denote the Fronsdal fields with frame indices by $\Phi^{A_1 \cdots A_S}$ and again employ an auxiliary Fock space description
\bea
\Phi = \sum_{s=0}^\infty \Phi_{A_1 \cdots A_S} \alpha^{A_1} \cdots \alpha^{A_S} |0\rangle
\eea
The creation $\alpha^{A}$ and annihilation  $\bar{\alpha}^A$ operators obey the commutator
\bea
[\bar{\alpha}^A, \alpha^B] = \eta^{AB}
\eea
where $A, B$ run over all frame field dimensions. We will also use the indices $(A)=(+, -, I)= (+, -, z, i)$. The double traceless condition is
\bea
(\bar{\alpha}^2)^2 \Phi\equiv (\bar{\alpha}\cdot\bar{\alpha})^2 \Phi =(\bar{\alpha}^A\bar{\alpha}_A)^2\Phi= 0
\eea
while the gauge transformation is
\bea
\Phi^\prime  = \Phi + \alpha^A D_A \Lambda
\eea
with a traceless ($\bar{\alpha}^2 \Lambda = 0$) gauge parameter. The AdS covariant derivative in frame field indices is
\bea
D_A \equiv \hat{\partial}_A + \frac{1}{2} \omega_A{}^{BC}\eta_{BD}\eta_{CE} M^{DE} \qquad M^{BC} = \alpha^B \bar{\alpha}^C - \alpha^C \bar{\alpha}^B
\eea
with $\hat{\partial}_A \equiv e_A^\mu \partial_\mu$ and $\omega_A{}^{BC}$ is the frame field spin connection for Poincar\'{e} AdS.  The equation of motion for the higher spin fields is
\bea
\left ( D^AD_A +\omega_A{}^{AB} D_B - s^2 + 2s + 2 - \alpha D \bar{\alpha}D + \frac{1}{2} (\alpha D)^2 \bar{\alpha}^2 - \alpha^2 \bar{\alpha}^2\right ) \Phi = 0
\eea
where we use the shorthand $\alpha D \equiv \alpha^A D_A$ and $\bar\alpha D \equiv \bar\alpha^A D_A$. In light-cone gauge $\Phi$ is traceless $\bar{\alpha}^2 \Phi = 0$ and after some work, the equations of motion become \cite{Metsaev:1999ui}
\bea
z^2 \partial^A \partial_A \left ( \frac{\Phi}{z} \right ) = 0 \label{HSeom}
\eea
This is the equation of motion obtained after fixing the gauge and solving the constraint associated with this gauge choice.

To determine the action of the conformal transformations on the higher spin gauge fields, we use the Lie derivative along the flow defined by the Killing vectors of the AdS isometries. The generators of these transformations do not, in general, preserve the light cone gauge choice. For this reason they must be supplemented by compensating gauge transformations\footnote{The Lorentz transformations are modified with a compensating gauge transformation in order that they preserve the light cone gauge condition. For this reason, tensors in the CFT do not map into tensors with the same indices in gravity. Indeed we will see that $-$, $x$ components of the current map into $X$, $Z$ components of the gauge fields. The details of this matching agrees perfectly with the GKPW mapping after transforming to the lightcone gauge \cite{Mintun:2014gua}.} which restore the gauge. This analysis has been carried out in detail in \cite{Metsaev:1999ui} and the complete set of generators are given in Section 3.8 of \cite{Metsaev:1999ui}.

As discussed above, only two components of the higher spin gauge field, at each spin, are physical and independent degrees of freedom. We will choose these two components to be $\Phi^{XX\cdots XX}$ and $\Phi^{XX\cdots XZ}$. Collect the complete set of physical and independent fields into a single field, with the help of an additional variable $\theta$ as follows
\bea
\Phi(X^+,X^-,X,Z,\theta)=\sum_{s=0}^\infty \left(\cos (2s\theta) {\Phi^{XX\cdots XX}\over Z}+\sin (2s\theta){\Phi^{XX\cdots XZ}\over Z}\right)\label{thetaexpansion}
\eea
For what follows it is again convenient to perform a Fourier transform to obtain
\bea
\Phi(X^+,P^+,X,Z,\theta)=\int \,dX^-\, e^{iP^+X^-} \Phi(X^+,X^-,X,\theta)
\eea
The equation of motion (\ref{HSeom}) becomes
\bea
i{\partial\over\partial X^+}\Phi(X^+,P^+,X,Z,\theta)=-{1\over 2P^+}\left({\partial^2\over\partial X^2}+{\partial^2\over\partial Z^2}\right)\Phi(X^+,P^+,X,Z,\theta)\label{HSGeom}
\eea

\subsection{Holography}\label{lightfrontholography}

The equal $x^+$ bilocal field $\eta(x^+,p_1^+,x_1,p_2^+,x_2)$ packages a scalar field of dimension $\Delta=1$ and the two independent components at each spin $2s$ of spinning conserved currents. It is a single field that is a function of 5 coordinates. The higher spin field $\Phi(X^+,P^+,X,Z,\theta)$ packages a bulk scalar as well as two physical and independent components of a spinning gauge field at each spin $2s$. It is a single field that is a function of 5 coordinates. Bilocal holography explicitly demonstrates the equality of these degrees of freedom by giving the identification between these two fields. This mapping is determined entirely by conformal symmetry. The action of the conformal group on the higher spin field $\Phi^{A_1\cdots A_{2s}}(X^+,X^-,X,Z)$ (and hence also on $\Phi(X^+,P^+,X,Z,\theta)$) follows from the analysis of  \cite{Metsaev:1999ui}, while the action of the conformal group on the bilocal is given by the coproduct of the usual free scalar field representation. A key observation, derived in \cite{deMelloKoch:2010wdf,deMelloKoch:2021cni}, is that the generators of these two representations are exactly mapped to each other through the identification of the coordinates
\bea
x_1&=& X+Z \tan \left(\frac{\theta }{2}\right)\qquad
x_2\,\,=\,\, X-Z \cot \left(\frac{\theta }{2}\right)\qquad x^+=X^+\cr
p_1^+&=& P^+ \cos ^2\left(\frac{\theta }{2}\right)\qquad\quad
p_2^+\,\,=\,\, P^+ \sin ^2\left(\frac{\theta }{2}\right)\label{mapcft2grav}
\eea
and the fields
\bea
\Phi  &=& 2\pi P^+\sin\theta\,\, \eta\label{identifyfields}
\eea
The inverse of (\ref{mapcft2grav}) is
\bea
X&=& \frac{p_1^+ x_1+p_2^+ x_2}{p_1^++p_2^+}\qquad
Z\,\,=\,\,\frac{\sqrt{p_1^+ p_2^+} (x_1-x_2)}{p_1^++p_2^+}\cr
P^+&=& p_1^++p_2^+\qquad\qquad
\theta\,\,=\,\,2 \tan ^{-1}\left(\sqrt{\frac{p_2^+}{p_1^+}}\right)\label{mapgrav2cft}
\eea
The basic claim of bilocal holography is that this mapping between the coordinates of the CFT and those of AdS$_4$, as well as the identification between the bilocal and the higher spin fields, provides a construction of the higher spin quantum gravity starting from the conformal field theory. This claim passes some highly non-trivial checks. 

At the most basic level, we should verify that we have obtained a valid bulk reconstruction. Under the identification (\ref{mapcft2grav}), the CFT equation of motion (\ref{CFTeom}) is mapped to the higher spin equation of motion (\ref{HSGeom}). To see this, start from the CFT equation of motion, which implies that
\bea
i\partial_+ \eta&=&-\left({1\over 2p_1^+}{\partial^2\over\partial x_1^2}+{1\over 2p_2^+}{\partial^2\over\partial x_2^2}\right)\eta
\eea
Since $\eta$ and $\Phi$ are proportional to each other, with the constant of proportionality (see equation (\ref{identifyfields}) above) independent of $x_1$ and $x_2$, we know that $\Phi$ obeys the same equation of motion. Thus, we have (the second equality below uses the chain rule as well as (\ref{mapgrav2cft}))
\bea
i\partial_+ \Phi&=&-\left({1\over 2p_1^+}{\partial^2\over\partial x_1^2}+{1\over 2p_2^+}{\partial^2\over\partial x_2^2}\right)\Phi\cr\cr
&=&-{1\over 2P^+}\left({\partial^2\over\partial X^2}+{\partial^2\over\partial Z^2}\right)\Phi
\eea
This proves that the complete tower of higher spin fields obey the correct equations of motion. Do these fields satisfy the correct boundary conditions? This entails studying the $Z\to 0$ limit of the bulk fields \cite{Banks:1998dd,Harlow:2011ke}. The usual GKPW dictionary \cite{Gubser:1998bc,Witten:1998qj} is formulated in de Donder gauge, where the solution to the higher spin equations of motion behaves as
\bea
\Phi^{M_1\cdots M_{2s}}\sim Z^{2-2s} A^{M_1\cdots M_{2s}} (X^+,X^-,X,Z=0)+Z^{2s+1}
B^{M_1\cdots M_{2s}}(X^+,X^-,X,Z=0)\nonumber
\eea
as $Z\to 0$. Here the indices $M_i$ take values $+,-,X$. Components of the gauge field with $k$ $Z$ polarizations behave as
\bea
\Phi{M_1\cdots M_{2s-k}Z\cdots Z}&\sim& Z^{2-2s-k}A^{M_1\cdots M_{2s-k}Z\cdots Z}(X^+_,X^-,X,Z=0) \cr\cr
&&\qquad+ Z^{2s+1+k}B^{M_1\cdots M_{2s-k}Z\cdots Z}(X^+,X^-,X,Z=0)
\eea
Picking up the leading term of the normalizable solution, and noting that (at leading order in $Z$) we have conservation and tracelessness of the identified tensors, we obtain the holographic dictionary
\bea
B^{M_1\cdots M_{2s}}(X^+,X^-,X,Z=0)&=&j_{(2s)}^{M_1\cdots M_{2s}}(X^+,X^-,X)\label{GKPW}
\eea
with $j_{(2s)}^{M_1\cdots M_{2s}}(X^+,X^-,X)$ the CFT primary. Our fields do not obey this boundary condition. Rather, as $Z\to 0$ we find
\bea
{\partial^{2s}\over \partial X^-{}^{2s}}\Phi_{2s}(X^+;X^-,X,0)
&=&16\pi{\cal N}\sum_{k=0}^{2s}
\frac{(-1)^k \partial_-^{2s-k}\phi^a(X^+,X^-,X)\partial_-^{k}\phi^a(X^+,X^-,X) }
{\Gamma \left(2s-k+\frac{1}{2}\right) \Gamma \left(k+\frac{1}{2}\right) k! (2s-k)!}\cr
&&\label{LC}
\eea
\bea
{\cal N}={(2s)!\over\Gamma\left(2s+{1\over 2}\right)}
\eea
which is easily proved \cite{deMelloKoch:2014vnt} by making use of the identity
\bea
(p_1^++p_2^+)^{2s}\, \cos \left(4s \tan^{-1}\sqrt{p_2^+\over p_1^+}\right)
={\cal N}
\sum_{k=0}^{2s}
\frac{(-1)^k (p_1^+)^{2s-k}(p_2^+)^k}
{\Gamma \left(2s-k+\frac{1}{2}\right) \Gamma \left(k+\frac{1}{2}\right) k! (2s-k)!}\label{usefulidentity}
\eea
This apparent discrepancy was resolved in \cite{Mintun:2014gua}. The basic point is that, as we have observed, the GKPW dictionary is obtained in the de Donder gauge while the bilocal mapping, is written in lightcone gauge. Under the change of gauge from de Donder to light cone gauge, the boundary condition (\ref{GKPW}) transforms into (\ref{LC}). This proves that the complete collection of fields in the higher spin gravity obey the correct equations of motion with the correct boundary condition.

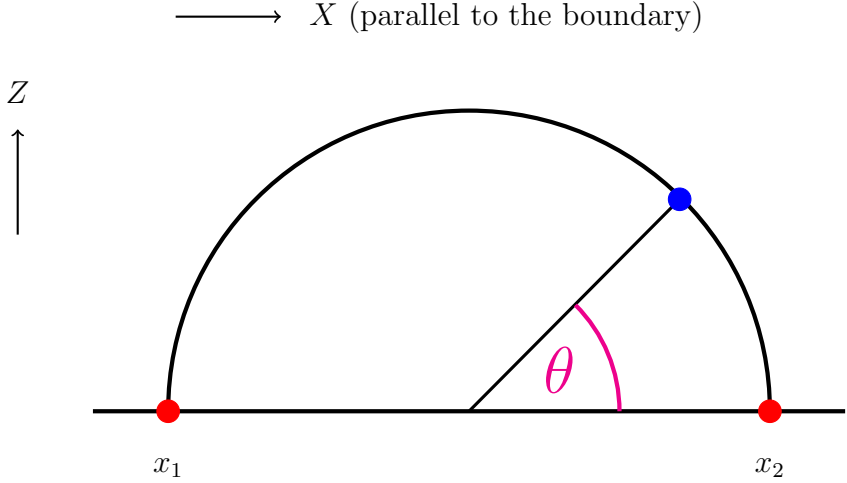
\begin{figure}[ht]%
\begin{center}
\begin{tikzpicture}
\draw[ultra thick] (1,0.75) -- (11,0.75);
\draw[ultra thick] (2,0.75) arc (180:0:4 and 4);
\filldraw[red] (2,0.75) circle (0.15);
\filldraw[red] (10,0.75) circle (0.15);
\draw node at (2,0) {$x_1$};
\draw node at (10,0) {$x_2$};
\draw[very thick](6,0.75) -- (8.8,3.57);
\filldraw[blue] (8.8,3.57) circle (0.15);
\draw[ultra thick,magenta] (8,0.75) arc (0:45:2);
\draw[magenta] node at (7.2,1.3) {{\huge $\theta$}};
\draw[->,thick] (2.1,6) -- (3.5,6);
\draw node at (6.5,6) {$X$ (parallel to the boundary)};
\draw[->,thick] (0,3.1) -- (0,4.5);
\draw node at (0,5) {$Z$};
\end{tikzpicture}
\caption{The horizontal direction, parametrized by $X$, is parallel to the boundary. The vertical direction, perpendicular to the boundary is parametrized by the emergent holographic coordinate $Z$. The semicircle centre is at ${x_1+x_2\over 2}$.}%
\label{definetheta}%
\end{center}
\end{figure}

Another property of the map that can be explored regards subregion duality: for every given CFT subregion ${\bf A}$ together with some code subspace, there exists a maximal bulk subregion ${\bf a}$ whose algebra of operators ${\cal A}_a$ acting on the code subspace can be encoded in the algebra of boundary operators ${\cal A}_A$. The region ${\bf a}$ is
the bulk region bounded by ${\bf A}$ and its Ryu-Takayanagi (RT) surface at leading order in the gravitational coupling $G_N$. We recall that the RT surface is the minimal area extremal surface homologous to the boundary region ${\bf A}$, whose area in Planck units gives the CFT entropy of the boundary region ${\bf A}$ to leading order in the $G_N$ expansion. The bulk region ${\bf a}$ is referred to as the entanglement wedge. From the point of view of the bilocal holographic map, a simple example of a CFT subregion ${\bf A}$ is to allow $x^-$ (and hence $p^+$) to be unrestricted, but to restrict $-{L\over 2}\le x\le {L\over 2}$. Consider a bilocal composed of two excitations that are described by wave packets tightly peaked about some position $x$ and some momentum $p^+$. Locate the first excitation at $x_1$ and $p_1^+$, and the second at $x_2$ and $p_2^+$. Where is the corresponding bulk excitation located? Using the mapping (\ref{mapgrav2cft}) it is simple to verify that
\bea
\left(X-{x_1+x_2\over 2}\right)^2+Z^2 =\left({x_1-x_2\over 2}\right)^2\label{ewreqn}
\eea
This locates the excitation dual to this bilocal on a semi-circle in the bulk. To localize the excitation to a definite bulk position, we need to localize on angle $\theta$ in Figure \ref{definetheta}.

\begin{figure}[ht]%
\begin{center}
\begin{tikzpicture}
\draw[ultra thick] (1,0.75) -- (11,0.75);
\filldraw[gray] (2,0.75) arc (180:0:4 and 4);
\draw[ultra thick] (2,0.75) arc (180:0:4 and 4);
\filldraw[red] (2,0.75) circle (0.15);
\filldraw[red] (10,0.75) circle (0.15);
\draw node at (2,0) {$x_1=-{L\over 2}$};
\draw node at (10,0) {$x_2={L\over 2}$};
\draw[thick] (3,0.75) arc (180:0:2.5 and 2.5);
\filldraw[blue] (3,0.75) circle (0.15);
\filldraw[blue] (8,0.75) circle (0.15);
\draw[thick] (6,0.75) arc (180:0:1.5 and 1.5);
\filldraw[blue] (6,0.75) circle (0.15);
\filldraw[blue] (9,0.75) circle (0.15);
\draw[->,thick] (2.1,6) -- (3.5,6);
\draw node at (6.5,6) {$X$ (parallel to the boundary)};
\draw[->,thick] (0,3.1) -- (0,4.5);
\draw node at (0,5) {$Z$};
\end{tikzpicture}
\caption{We consider a subregion ${\bf A}$ of the CFT defined by taking $-{L\over 2}<x<{L\over 2}$. Bilocal operators belonging to this CFT subregion can be used to construct bulk operators lying in the grey region, which is the corresponding entanglement wedge {\bf a}. The bilocal with excitations located at $x_1=-{L\over 2}$ and $x_2={L\over 2}$ corresponds to a bulk excitation that can be localized on the boundary of the entanglement wedge. By considering bilocal fields with their excitations (shown as blue circles in the figure) inside the subregion ${\bf A}$, we construct bulk operators lying inside the entanglement wedge.}%
\label{EWRfig}%
\end{center}
\end{figure}
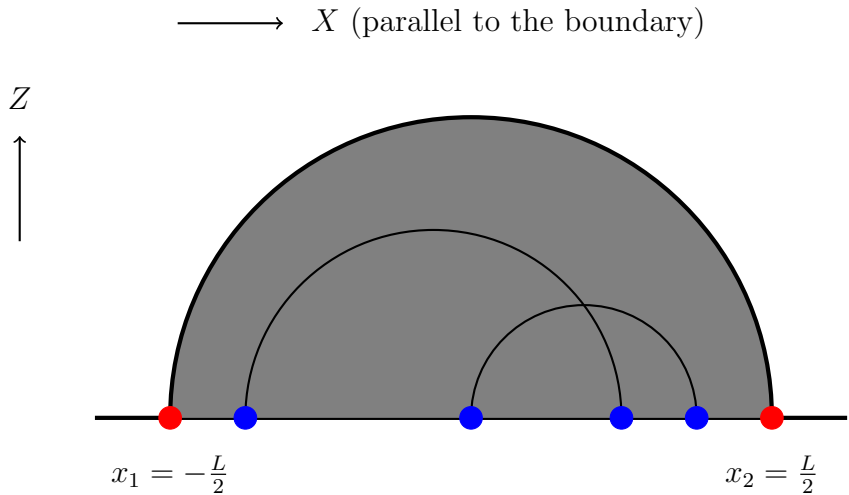

\noindent
Simple trigonometry shows that
\bea
\tan\theta ={Z\over X-{x_1+x_2\over 2}}={\sqrt{p_1^+p_2^+}\over p_1^++p_2^+}
\eea
and further, that this angle $\theta$ is exactly the angle appearing in (\ref{mapcft2grav}) and (\ref{mapgrav2cft}), which justifies its name. Thus, it is by localizing the CFT excitations in both $x$ and $p^+$ that allows us to localize an excitation in the bulk. This tells us that the collection of bilocals with both excitations located in the CFT subregion ${\bf A}$ explore the region of the bulk defined by
\bea
  X^2+Z^2 \le \left({L\over 2}\right)^2\qquad {\rm and\,\,any}\quad X^-
\eea
See Figure \ref{EWRfig} for an illustration. The boundary of this region is the union of the subregion in the CFT and an extremal surface in the bulk, so that we have naturally reproduced the statement of entanglement wedge reconstruction \cite{deMelloKoch:2021cni}.

The behaviour of the angle $\theta$ as a function of $p_1^+$ and $p_2^+$ is interesting to explore. $\theta$ positions us on the semicircle as shown in Figure \ref{definetheta}. At $p_1^+=p_2^+$ we have $\theta={\pi\over 2}$ so that the bulk excitation is as deep in the bulk as possible. When one of the $p^+$ momenta is very small compared to the other, the bilocal is dominated by the large $p^+$ excitation and becomes point like. In this case $\theta\sim 0$ or $\theta\sim\pi$ (depending on which particle dominates the bilocal) and the bulk excitation is again located close to the boundary.

There is one more point worth discussing: information localizes in bilocal holography exactly as expected from a theory of quantum gravity. As a consequence of the entanglement structure of the quantum vacuum and the Gauss Law of gravity, \cite{Laddha:2020kvp,Chowdhury:2020hse,Raju:2020smc,Raju:2021lwh,SuvratYouTube} have argued for the principle of the holography of information which states that ``In a theory of quantum gravity, a copy of all the information available on a Cauchy slice is also available near the boundary of the Cauchy slice. This redundancy in description is already visible in the low-energy theory.'' From the mapping (\ref{mapgrav2cft}) it is clear that single trace primaries and their descendants\footnote{For which both fields in the bilocal are at the same spatial position so that $Z=0$.} in the CFT map into the region at the boundary of AdS$_4$, while bilocals that are deep in the bulk have the two fields in the bilocal well separated. The holography of information is then implied by the operator product expansion of the CFT \cite{deMelloKoch:2022sul} which expresses a product of separated operators as a convergent sum of the local single trace primary operators and their descendants. This assumes that the single trace primaries and their descendants are complete in the sense that they generate all local operators when they are multiplied. This is reasonable since they are dual to the fields appearing in the dual gravity and hence they should generate the complete gravity Fock space.

Thus, bilocal holography explains the origin of the extra holographic dimension\footnote{It is determined by the separation between the two fields in the bilocal - see the formula for $Z$ in (\ref{mapgrav2cft}). This was one of the important conclusions reached in \cite{deMelloKoch:2010wdf}.}, it gives rise to fields that obey local field equations with the correct boundary conditions in this higher dimensionsal spacetime and we have some evidence that it localizes information exactly as is expected in a theory of gravity.

\subsection{Comments on the holographic map}\label{HMapComments}

We have performed our analysis in the light cone gauge motivated by the fact that this gauge is particularly convenient for performing a total gauge fixing that reduces the theory to its physical degrees of freedom. It is not easy to perform a complete gauge fixing in other gauges and hence one suspects that it will not be easy to repeat our argument in this case. With this observation in mind, it is worth studying the lightcone map to see what general lessons we can extract, since these may provide a more efficient route to constructing the map in other gauges. This is the goal of this section.

It is possible to give the map as a change of coordinates in momentum space. The first two entries of the map simply reflect the fact that the boundary CFT and the bulk gravity share the same translation invariance, so that we can equate the conserved charges of these symmetries
\bea
P^+&=&p_1^++p_2^+\qquad\qquad\qquad P\,\,=\,\,p_1+p_2
\eea
In momentum space, the equations of motion in the CFT imply the equations of motion of the higher spin gravity if we choose
\bea
P^z&=&\sqrt{p_2^+\over p_1^+}p_1-\sqrt{p_1^+\over p_2^+}p_2
\eea
Finally, the angle $\theta$ summarizes the structure of the spinning CFT primary conserved currents. They play a role only at $Z=0$, where the boundary conditions requiring that the bulk field correctly reproduces the CFT operators, are imposed. This determines
\bea
\theta&=&2 {\rm arctan}\sqrt{p_2^+\over p_1^+}
\eea
To see this, it is useful to recall equations (\ref{LC}) and (\ref{usefulidentity}).

Now consider the position space version of the map. The formulas for $X$ and $X^-$ look like centre of mass coordinates, for the pair of excitations in the bilocal
\bea
X&=&{p_1^+ x_1 +p_2^+ x_2\over p_1^++p_2^+}\qquad
X^-\,\,=\,\,{p_1^+ x_1^- +p_2^+ x_2^-\over p_1^++p_2^+}
\eea
Note that $p_1^+$ and $p_2^+$ play the role of masses, which is natural from the point of view of light front kinematics. The centre of mass coordinate is the obvious formula to associate to an extended body. The $Z$ coordinate is now determined to be
\bea
Z&=&{\sqrt{p_1^+p_2^+}\over p_1^++p_2^+}(x_1-x_2)\label{holcoord}
\eea
by the requirement that we obtain the correct entanglement wedge for the CFT subregion described by $-{L\over 2}\le x\le {L\over 2}$. The requirement that we obtain the correct entanglement wedge is given in equation (\ref{ewreqn}). The resulting formula for the holographic coordinate $Z$, given in (\ref{holcoord}), has a very important property: it locates bulk excitations dual to the single trace primaries (which have $x_1=x_2$) in an arbitrarily small neighbourhood of the boundary, while bilocals with well separated fields map to bulk excitations located deep in the bulk. Using the OPE we can express bilocals (and their products) in terms of single trace primaries (and their products), so that our formula for $Z$ is perfectly consistent with the principle of the holography of information. In this way, collective field theory is providing a geometrization of the space of CFT operators in a manner that is perfectly consistent with how we expect information to localize in a theory of quantum gravity. Finally, the angle $\theta$ again summarizes the structure of the CFT primary operators, and this structure is again only relevant at $Z=0$ where the boundary condition related the boundary behaviour of bulk gravity fields to the operators in the CFT. From Figure \ref{definetheta} it is clear that $\theta$ is a ``local angle'' defined with respect to an origin defined locally by the bilocal. Looking at equation (\ref{thetaexpansion}) we see that $\theta$ is used as a book keeping device to collect the different spin states. In this sense it is not too different from a polarization. In the case of light for example, polarization is defined as transverse to the direction of motion, i.e. it too is a defined locally with respect to the photons direction of propagation.

As a final comment, on the gravity side to reduce to physical degrees of freedom, we have to choose a gauge. On the CFT side this reduction amounted to working with the equal $x^+$ bilocal. Had we chosen a temporal gauge (for example), we would be eliminating temporal polarizations, in which case we would need to consider the equal time $t$ bilocal. This illustrates that the choice of which bilocal is used in CFT, is closely related to the gauge choice in the dual gravity.

\section{Covariant Bilocal Holography}\label{covariant}

In the present discussion, the term ``covariant" refers to the preservation of the boundary Poincar\'{e} symmetry. In the light front approach to bilocal holography we identified two independent components of the current at each spin. In the higher spin gravity this entailed fixing light cone gauge and its associated constraint, leaving only components of the gauge field with $X$, $Z$ polarizations. This obviously does not preserve the boundary Poincar\'{e} symmetry. In the CFT description, the reduction to independent components requires solving the current conservation equation to eliminate redundant components of the current and this step necessarily breaks the boundary Poincar\'{e} invariance. To maintain this symmetry, we will not solve the current conservation equation, at the cost of working with a larger set of redundant variables. In the dual gravity we will employ a modified de Donder gauge \cite{Metsaev:2008ks} (see also \cite{Metsaev:2009hp,Metsaev:2011uy,Metsaev:2013wza}) which preserves the boundary Poincar\'{e} invariance. The goal is then to establish a correspondence between this larger set of variables and a redundant set of variables in higher spin gravity. Fortunately, a well-executed and insightful paper by Metsaev \cite{Metsaev:2008fs} has already done this, providing valuable insights that we will use heavily in this section. Once the correspondence is established, we can then solve the current conservation equation and again reduce both sides to physical and independent degrees of freedom. We will follow this route to obtain the map for the equal time bilocal theory.

The construction of \cite{Metsaev:2008fs} introduces extra fields into the CFT description and imposes a constraint on this bigger set of fields. There is a redundancy in this extra field description, which can be expressed as a ``gauge symmetry''. There is an action of the conformal group on this larger collection of fields. In the dual higher spin gravity, the modified de Donder gauge is used. This gauge has the attractive feature that it leads to decoupled equations of motion, at every spin, that can be explicitly solved. In this gauge, on shell, there is a residual gauge symmetry. The analysis of \cite{Metsaev:2008fs} demonstrates that the modified de Donder gauge condition and the residual gauge symmetry of the higher spin gravity matches the constraint imposed and the ``gauge symmetry'' of the CFT. The action of the so(2,$d$) symmetry generators on the two sides also match. 

Once again, following \cite{Metsaev:2008fs}, it is convenient to assemble fields into ket vectors $|\phi\rangle$. For this purpose, we use the same oscillators $(\alpha^a,\alpha^Z)$ for the gravity and CFT descriptions. Here the index $a$ runs over the coordinate labels in CFT. The oscillator $\alpha^Z$ is a book keeping device in the CFT. In the dual gravity it plays a more physical role since it is associated to the holographic dimension $Z$. The oscillator commutators are
\bea
[\bar{\alpha}^a,\alpha^b]=\eta^{ab}\qquad [\bar{\alpha}^Z,\alpha^Z]=1\qquad
[\bar{\alpha}^a,\alpha^Z]=0=[\bar{\alpha}^Z,\alpha^a]
\eea
The so(2,$d$) generators can then be written as
\bea 
 \delta_{\hat{G}}|\phi\rangle  = \hat{G} |\phi\rangle 
\eea
where $\hat{G}$ is a differential operator, given by
\bea
P^a &=& \partial^a \qquad J^{ab}\,\,=\,\,x^a\partial^b-x^b\partial^a+M^{ab}\qquad
M^{ab} \,\,\equiv\,\, \alpha^a \bar\alpha^b - \alpha^b\bar\alpha^a\cr\cr
D &=& x\cdot\partial  + \Delta\qquad K^a\,\,=\,\, -\frac{1}{2}x^2\partial^a + x^a D
+ M^{ab}x^b + R^a
\eea
Thus, the representation is determined by giving $R^a$ and $\Delta$. This form of the generators is valid in both the CFT and in the higher spin gravity, once the correct $\Delta$ and $R^a$ are determined.

In Sections \ref{covariantcft} and \ref{covariantgravity} we review the construction of \cite{Metsaev:2008fs} and then use it, in Section \ref{covariantmapping} to determine the map of covariant bilocal holography. This gives the holographic map for the equal time bilocal field. Once again we are able to verify that the mapping gives a valid bulk reconstruction, it realizes the holography of information and it provides a simple explanation of entanglement wedge reconstruction.

\subsection{Covariant CFT}\label{covariantcft}

The field content is summarized, using the oscillator language, as follows
\bea
|\phi\rangle\equiv\sum_{s'=0}^s\alpha_Z^{s-s'}|j_{s'}\rangle\qquad
|j_{s'}\rangle\equiv\frac{\alpha_{a_1}\ldots\alpha_{a_{s'}}}{s'!\sqrt{(s-s')!}}\, j_{s'}^{a_1\ldots a_{s'}} |0\rangle
\eea
where $j_{s'}^{a_1\ldots a_{s'}}$ is a rank-$s'$ tensor, traceless for $s'=2,3$ and double-traceless for $s'\geq 4$
\bea
\eta_{ab}\, j_{2}^{ab}=\eta_{ab}\, j_{3}^{aba_3}=0 \qquad\qquad
\eta_{ab}\,\eta_{cd}\,j_{s'}^{abcda_5\ldots a_{s'}}=0 \qquad s'=4,5,\ldots,s
\eea
We use the notation $j_{s'}^{a_1\ldots a_{s'}}$ for the fields since they represent the conserved currents of CFT. These fields have dimension $\Delta(j_{s'}^{a_1\ldots a_{s'}})=s'+d-2$. $|\phi\rangle$ is a degree-$s$ polynomial in $\alpha^a,\alpha^Z$ and $|j_{s'}\rangle$ is a degree-$s'$ homogeneous polynomial in $\alpha^a$
\beq 
(N_\alpha+N_Z)|\phi\rangle=s|\phi\rangle\qquad\qquad N_\alpha|j_{s'}\rangle=s'|j_{s'}\rangle
\eeq
where we have introduced the number operators $N_\alpha=\eta_{ab}\alpha^a\bar\alpha^b$ and $N_Z=\alpha^Z\bar\alpha^Z$. The double-tracelessness constraint is written as
\bea 
(\bar{\alpha}\cdot\bar{\alpha})^2 |\phi\rangle\equiv (\eta_{ab}\bar{\alpha}^a\bar{\alpha}^b)^2 |\phi\rangle  = 0 
\eea
This defines the $\cdot$ notation which indicates a contraction over the CFT directions i.e. the $Z$ index is not summed. This is in contrast to Section \ref{reducehsg} where the dot indicated a contraction, including $Z$.

This description employs more fields than required to describe a symmetric, traceless and conserved spinning current of spin $s$. Consequently, the description is redundant and so it is possible to impose a constraint. In the next section it will become clear that the constraint we impose in the CFT matches the modified de Donder gauge condition imposed in the higher spin gravity. The CFT constraint can be written as
\bea
\bar{C}_{CFT}|\phi\rangle=0\label{CFTconstraint}
\eea
where the operator $\bar{C}_{CFT}$ is given by
\bea 
\bar{C}_{CFT}=\bar{\alpha}\cdot\partial-\half\alpha\cdot\partial\,\bar\alpha\cdot\bar\alpha +\half \alpha^Z\,\widetilde{e}_1\,\bar\alpha\cdot\bar\alpha+\widetilde{e}_1\,\bar\alpha^Z \,\Box\,\Pi
\eea
where
\bea
\Pi = 1 -\alpha\cdot\alpha\frac{1}{2(2N_\alpha +d)}\bar\alpha\cdot\bar\alpha\qquad
\qquad \widetilde{e}_1=\sqrt{\frac{2s+d-4-N_Z}{2s+d-4-2N_Z}}
\eea
The operator $\widetilde{e}_1$ evaluates to an $s,s'$ dependent number for each term summed in $|\phi\rangle$. Since $j_{s'}^{a_1\ldots a_{s'}}$ is double traceless, it can be uniquely decomposed into the sum of a traceless rank $s'$ tensor and a traceless rank $s'-2$ tensor. The operator $\Pi$ is a projector, which projects onto the traceless rank $s'$ piece.

There are local transformations of the fields that leave the constraint (\ref{CFTconstraint}) invariant. Following \cite{Metsaev:2008fs} we refer to these local transformations as ``gauge symmetries'' of the CFT. These local transformations match the on shell residual gauge transformations that remain after fixing the modified de Donder gauge in the higher spin gravity. The parameters of these local transformations are $\xi_{s'}^{a_1\ldots a_{s'}}$, $s'=0,1,\ldots, s-1$. They can be collected into a vector $|\xi\rangle$ defined by
\bea
  |\xi\rangle\equiv\sum_{s'=0}^{s-1}\alpha_Z^{s-1-s'}|\xi_{s'}\rangle \qquad
  |\xi_{s'}\rangle\equiv\frac{\alpha_{a_1}\ldots\alpha_{a_{s'}}}{s'!\sqrt{(s -1 - s')!}} \,   
  \xi_{s'}^{a_1\ldots a_{s'}} |0\rangle
\eea
so that
\bea
  (N_\alpha+N_Z)|\xi\rangle=(s-1)|\xi\rangle\qquad 
   N_\alpha|\xi_{s'}\rangle=s'|\xi_{s'}\rangle
\eea
These parameters are totally symmetric tensor fields, have dimension $\Delta(\xi_{s'}^{a_1\ldots a_{s'}})=s'+d-3$ and are traceless
\bea 
  \bar\alpha^2 |\xi\rangle\equiv \eta_{ab}\,\bar\alpha^a\bar\alpha^b |\xi\rangle = 0 
\eea
The ``gauge transformations'' that leaves the constraint (\ref{CFTconstraint}) invariant are
\bea
\delta_{CFT}|\phi\rangle &=& (\alpha\cdot\partial-\alpha^Z\widetilde{e}_1+\frac{1}{2s+d-6 -2N_Z}\,\widetilde{e}_1\, \bar\alpha^Z \alpha\cdot\alpha\,\Box)|\xi\rangle\label{CFTGauge}
\eea
The representation of the conformal group on these fields is defined by $\Delta=s+d-2-N_Z$ and
\beq
R^a&=&\bar{r}\Bigl(\alpha^a-\alpha\cdot\alpha\frac{1}{2N_\alpha + d-2}\bar\alpha^a+\alpha\cdot\alpha
\frac{2}{(2N_\alpha+d-2)(2N_\alpha+d)}(\bar\alpha^a-\frac{1}{2}\alpha^a \bar\alpha\cdot\bar\alpha)\Bigr)\cr\cr
\bar{r}&\equiv&-\Bigl((2s+d-4-N_Z)(2s+d-4-2N_Z)\Bigr)^{1/2} \bar\alpha^Z
\eeq

It is possible to ``choose a gauge'' that recovers the standard CFT description of the current. Using a compact notation $j_{s'}\sim j_{s'}^{a_1\ldots a_{s'}}$, $\xi_{s'}\sim\xi_{s'}^{a_1\ldots a_{s'}}$, $\partial\sim\partial^a$ and $\eta\sim\eta^{ab}$ the gauge transformation (\ref{CFTGauge}) is
\bea 
\delta j_{s'}&\sim&\partial \xi_{s'-1}+\xi_{s'}+\eta\,\Box\,\xi_{s'-2}\qquad s' = 2,3,\ldots,s\cr\cr
\delta j_1&\sim&\partial\xi_0+\xi_1\qquad \delta j_0\,\,\sim\,\, \xi_0\label{GT}
\eea
and $\xi_s\equiv 0$. The currents $j_{s'}$ with $s'\geq 2 $ decompose as
\bea 
j_{s'}&=&j_{s'}^{\rm T} \oplus j_{s'-2}^{\rm TT}\,, \qquad s'= 2,3,\ldots,s 
\eea
where $j_{s'}^{\rm T}$ and $j_{s'-2}^{\rm TT}$ are rank-$s'$ and rank-$(s'-2)$ traceless tensors.  From (\ref{GT}), it is clear that we can use $\xi_0$ to set $j_0$ to zero, $\xi_1$ to set $j_1$ to zero and $\xi_{s'}$ to set $j^T_{s'}$ to zero for $s'= 2,3,\ldots,s-1$. These further ``gauge conditions'' can be written as
\bea
\Pi |j_{s'}\rangle = 0\qquad s'= 0,1,\ldots ,s-1 \qquad{\rm or}\qquad
\bar\alpha^Z\,\Pi|\phi\rangle=0
\eea
or equivalently
\bea 
|j_{s'}\rangle = \alpha\cdot\alpha \frac{1}{2(2N_\alpha +d)}\bar\alpha\cdot\bar\alpha |j_{s'}\rangle,\ \ \ s'=0,1,\ldots,s-1\label{usefulident}
\eea
After making this choice, the constraint (\ref{CFTconstraint}) implies that
\bea
(\bar{\alpha}\cdot\partial - \half \alpha\cdot\partial\,\bar\alpha\cdot\bar\alpha)|j_{s'}\rangle\! + \! \half \,\widetilde{e}_1|_{N_Z=s-s'-1}\,\bar\alpha\cdot\bar\alpha |j_{s'+1}\rangle\! =\! 0\qquad s'= 0,1,\ldots ,s
\eea
which can be expressed as
\bea
\frac{2N_\alpha + d-4}{2N_\alpha + d-2} \Bigl(\alpha\cdot\partial  - \alpha^2
\frac{1}{2N_\alpha + d}\bar{\alpha}\cdot\partial\Bigr)\bar\alpha\cdot\bar\alpha|j_{s'}\rangle+ \! \half \,\widetilde{e}_1|_{N_Z=s-s'-1}\,\bar\alpha\cdot\bar\alpha |j_{s'+1}\rangle\! =\! 0
\label{foriterat}
\eeq
when $s'= 0,1,2,\ldots ,s-1$, and for $s'=s$ we have
\bea
\Bigl(\alpha\cdot\partial-\frac{1}{2}\alpha\cdot\partial\bar\alpha^2\Bigr)|j_s\rangle=0
\label{LastConst}
\eea
Now, (\ref{foriterat}) implies that if $\bar\alpha\cdot\bar\alpha |j_{s'}\rangle=0$, then $\bar\alpha\cdot\bar\alpha |j_{s'+1}\rangle=0$. Since $|j_0\rangle = 0$, $|j_1\rangle=0$ we thus obtain
\bea 
\bar\alpha^2|j_{s'}\rangle=0\qquad s'=0,1,\ldots, s
\eea
so that by (\ref{usefulident}) we obtain
\bea
|j_{s'}\rangle=0\,,\qquad s'=0,1,\ldots, s-1
\eea
In the end we are left with the one spin-$s$ traceless current $|j_{s}\rangle$ which, by (\ref{LastConst}) is conserved
\bea 
\bar{\alpha}\cdot\partial|j_{s}\rangle = 0
\eea
Thus, we have recovered the usual description of the spinning current $|j_{(s)}\rangle$ as a totally symmetric and traceless conserved spin $s$ field
\bea
|\phi\rangle = |j_s\rangle\qquad
|j_{s}\rangle =\frac{\alpha^{a_1}\ldots\alpha^{a_s}}{s!}\, j_{s}^{a_1\ldots a_{s}} |0\rangle
\eea
Acting on this state we have
\bea
\Delta |\phi\rangle =(s+d-2)|\phi\rangle\qquad\qquad R^a|\phi\rangle =0
\eea
so that we recover the standard so(2,$d$) generators. We could, if we like, reduce to physical degrees of freedom, by eliminating the $0$ polarizations with the current conservation equation and solving the traceless and symmetric conditions leaving two independent components. The techniques needed to carry this out in complete detail are described in \cite{deMelloKoch:2014vnt}. We will not need the details of this reduction.

\subsection{Covariant Higher Spin Gravity}\label{covariantgravity}

Following \cite{Metsaev:2008fs}, a massless spin-$s$ field in AdS$_{d+1}$ spacetime is described by a scalar, a vector, and totally symmetric tensor fields $\phi_{s'}^{a_1\ldots a_{s'}}$ for $s'=0,1,\ldots,s$. Collect these fields into a ket
\bea
|\phi^{(s)}\rangle\equiv\sum_{s'=0}^s\alpha_Z^{s-s'}|\phi_{s'}\rangle\qquad |\phi_{s'}\rangle \equiv\frac{\alpha_{a_1} \ldots \alpha_{a_{s'}}}{s'!\sqrt{(s - s')!}}\,\phi_{s'}^{a_1\ldots a_{s'}}|0\rangle
\eea
which obeys
\bea 
(N_\alpha+N_Z)|\phi^{(s)}\rangle=s|\phi^{(s)}\rangle\qquad N_\alpha|\phi_{s'}\rangle=s'|\phi_{s'}\rangle 
\eea
The fields $\phi_{s'}^{a_1\ldots a_{s'}}$ with $s'>3$ are double-traceless which can be expressed as
\bea
(\bar{\alpha}\cdot\bar{\alpha})^2 |\phi^{(s)}\rangle \equiv (\eta_{ab}\,\bar{\alpha}^a\bar{\alpha}^b)^2 |\phi^{(s)}\rangle  =0
\eea
The modified de Donder gauge condition is
\bea 
\bar{C}_{AdS}|\phi^{(s)}\rangle  =  0 \label{mdD}
\eea
where $\bar{C}_{AdS}$ is given by
\bea 
\bar{C}_{AdS}&\equiv&\bar{\alpha}\cdot\partial-\half\alpha\cdot\partial\bar\alpha\cdot\bar\alpha-\half \alpha^Z \widetilde{e}_1\Bigl(\partial_Z+\frac{2s+d-5-2N_Z}{2Z}\Bigr) \bar\alpha\cdot\bar\alpha\cr\cr 
&&+\Bigl(\partial_Z - \frac{2s + d -5-2N_Z}{2Z}\Bigr) \widetilde{e}_1 \bar\alpha^Z\Pi
\eea
The importance of this gauge is because we obtain decoupled equations of motion \cite{Metsaev:2008ks}
\bea
\Bigl(\Box+\partial_Z^2-\frac{1}{Z^2}\left(\nu^2-\frac{1}{4}\right)\Bigr)|\phi^{(s)}\rangle=0\qquad
\nu \equiv s + \frac{d-4}{2} - N_Z\label{HSeom}
\eea
This is an important simplification since, as we will see below, these equations are easily solved and the explicit form of the solution in indispensable in matching to the CFT. The gauge condition and equations of motion are invariant under the following residual on-shell gauge transformation
\bea
\delta_{AdS}|\phi^{(s)}\rangle&=&\Bigg(\alpha\cdot\partial+\alpha^z\widetilde{e}_1\Bigl(\partial_Z+\frac{2s+d-5-2N_Z}{2Z}\Bigr)\cr\cr
&&+\frac{\alpha\cdot\alpha}{2s+d-6-2N_Z}\Bigl(\partial_Z-\frac{2s+d-5-2N_Z}{2Z}\Bigr) \widetilde{e}_1\bar\alpha^Z\Bigg)|\xi^{(s-1)}\rangle\label{AdSGT}
\eea
where the ket $|\xi^{(s-1)}\rangle$ is
\bea
|\xi^{(s-1)}\rangle \equiv \sum_{s'=0}^{s-1}\alpha_Z^{s-1-s'}|\xi_{s'}\rangle\qquad
|\xi_{s'}\rangle \equiv\frac{\alpha_{a_1} \ldots \alpha_{a_{s'}}}{s'!\sqrt{(s -1 - s')!}}\xi_{s'}^{a_1\ldots a_{s'}} |0\rangle 
\eea
The transformation parameters are traceless $\bar\alpha\cdot\bar\alpha|\xi^{(s-1)}\rangle =0$ and obey the equations of motion
\bea
\Bigl(\Box+\partial_Z^2-\frac{1}{Z^2}\left(\nu^2-\frac{1}{4}\right)\Bigr)|\xi^{(s-1)}\rangle =0
\eea
Notice that the number of currents $j_{s'}^{a_1\cdots a_{s'}}$ employed in the covariant CFT description of the spinning current matches the number of fields $\phi_{s'}^{a_1\cdots a_{s'}}$ used in this section. Further, the number of ``gauge'' transformation parameters in the CFT ``gauge symmetry'' description matches the number of residual gauge transformation parameters in the AdS$_{d+1}$ higher spin gravity, in modified de Donder gauge.

To go further we need the explicit solutions to the equations of motion. The normalizable and non-normalizable solutions are\footnote{In Section \ref{lightfront} our notation used capital letters for the coordinates of AdS$_4$ and little letter for CFT$_3$. In this section we use $x^a$ for CFT$_3$ and $Z,x^a$ for AdS$_4$. The identification between CFT coordinates and some of the AdS$_4$ coordinates is because the boundary Poincar\'{e} symmetry is preserved. We will revert to capital letters for AdS$_4$ coordinates in the next section.}
\bea 
|\phi^{(s)}(x,Z)\rangle=U_\nu (-)^{N_Z}|\phi^{(s)}(x)\rangle \label{EOMsoln}
\eea
where this solution depends on an arbitrary ket $|\phi^{(s)}(x)\rangle$ that depends only on $x^a$ and
\bea
U_\nu&\equiv&\sqrt{qZ}J_\nu (qZ)q^{-\nu-\half}\qquad q^2 = \Box
\eea
The asymptotic behaviour of our solutions is
\bea
|\phi^{(s)}(x,Z)\rangle\ \stackrel{Z\rightarrow 0}{\longrightarrow}\ Z^{\nu+\half}|\phi^{(s)}(x)\rangle=Z^s|\phi^{(s)}(x)\rangle\label{boundcond}
\eea

The representation of so(2,$d$) in this gauge requires some care. The modified de Donder gauge condition is invariant under both Poincar\'{e} and dilatation symmetries. It is not however invariant under transformations generated by the special conformal transformations. This can be corrected by supplementing $K^a$ with a compensating gauge transformation. The result of this analysis gives a representation of so(2,$d$) with
\bea
\Delta &=& Z\partial_Z +{1\over 2}(d-1)\qquad R^a\,\,=\,\,R^a_{(0)}+R^a_{(1)}
+R_{\rm comp}\label{AdSRep}
\eea
\bea
R^a_{(0)}&=&Z\left(\alpha^a-\alpha\cdot\alpha\frac{1}{2N_\alpha+d-2}\bar\alpha^a\right)\widetilde{e}_1\bar{\alpha}^Z-Z\alpha^Z\widetilde{e}_1\bar{\alpha}^a
\eea
\bea
R^a_{(1)}&=&-{1\over 2}Z^2\partial^a\qquad R_{\rm comp}|\phi^{(s)}(x,Z)\rangle\,\,=\,\,\delta_{\xi^{K^a}}|\phi^{(s)}(x,Z)\rangle
\eea
\bea
|\xi^{K^a}(x,Z)\rangle&=&ZU_{\nu+1}\left(\bar{\alpha}^a-\half \alpha^a\bar{\alpha}\cdot\bar\alpha\right)(-1)^{N_Z}|\phi^{(s)}(x)\rangle
\eea

Metsaev now provides a remarkable and explicit identification of the degrees of freedom in the CFT and in the AdS higher spin gravity. The proposal identifies the ket vector $|\phi^{(s)}(x)\rangle$ appearing on the RHS of (\ref{EOMsoln}) with the ket $|\phi\rangle$ of the previous subsection. There are three compelling pieces of evidence for this:
\begin{itemize}
\item[i.] The residual gauge transformations of AdS higher spin gravity reproduce the local symmetry of the CFT in the sense that
\bea
\delta_{AdS}|\phi^{(s)}(x,Z)\rangle&=& U_\nu (-)^{N_Z}\delta_{CFT}|\phi^{(s)}(x)\rangle
\eea

\item[ii.] The modified de Donder gauge condition of AdS higher spin gravity reproduces the constraint imposed in the CFT in the sense that
\bea
\bar{C}_{AdS}|\phi^{(s)}(x,Z)\rangle&=& U_\nu (-)^{N_Z}\bar{C}_{CFT}|\phi^{(s)}(x)\rangle
\eea

\item[iii.] The global so$(d,2)$ bulk symmetries of the massless spin-$s$ modes in AdS$_{d+1}$  become the global so$(d,2)$ boundary conformal symmetries ($G\in$so(2,$d$))
\bea
G_{AdS}|\phi^{(s)}(x,Z)\rangle&=& U_\nu (-)^{N_Z}G_{CFT}|\phi^{(s)}(x)\rangle
\eea
The proof of these statements uses the following useful identities
\bea
  \left(\partial_Z +\frac{\nu -\half}{Z}\right)U_\nu &=& U_{\nu-1}\qquad
  \left(\partial_Z-\frac{\nu + \half}{Z}\right)U_\nu\,\,=\,\,U_{\nu+1}(-\Box)\cr\cr
  \left(\partial_Z+\frac{\nu-\half}{Z}\right)U_{-\nu}&=&U_{-\nu+1}(-\Box )\qquad
  \left(\partial_Z-\frac{\nu +\half}{Z}\right)U_{-\nu}\,\,=\,\,U_{-\nu-1}
\eea
which are proved starting from the following Bessel function identities
\bea 
\left(\partial_Z + \frac{\nu}{Z}\right) J_{\nu}(Z) = J_{\nu-1}(Z)\qquad
\left(\partial_Z - \frac{\nu}{Z}\right) J_{\nu}(Z) = - J_{\nu+1}(Z)
\eea
\end{itemize}

In the conformal field theory we performed a ``gauge fixing'' that recovered the standard CFT description in terms of a traceless, symmetric and conserved spin $s$ current. We now consider the same gauge fixing in the AdS higher spin gravity. Following the decomposition used in the CFT, we decompose the fields $\phi_{s'}$ with $s'\geq 2 $ as follows
\bea 
\phi_{s'}&=&\phi_{s'}^{\rm T} \oplus \phi_{s'-2}^{\rm TT} \qquad\qquad s'= 2,3,\ldots,s 
\eea
Concretely, the decomposition is
\bea 
\alpha_{a_1}\cdots\alpha_{a_{s'}}\phi_{s'}^{a_1\cdots a_{s'}}&=&\alpha_{a_1}\cdots\alpha_{a_{s'}}((\phi_{s'}^{\rm T})^{a_1\cdots a_{s'}}+\eta^{a_1a_2}(\phi_{s'-2}^{\rm TT})^{a_3\cdots a_{s'}})
\eea
with both fields appearing, traceless
\bea
\eta_{ab}(\phi_{s'}^{\rm T})^{aba_3\cdots a_{s'}}\,\,=\,\,0\,\,=\,\,\eta_{ab}(\phi_{s'-2}^{\rm TT})^{aba_5\cdots a_{s'}}
\eea
The degree $s'$ piece in $\alpha^a$ of the gauge transformation law (\ref{AdSGT}) is given by
\bea
&&\alpha_{a_1}\cdots\alpha_{a_{s'}}((\delta_{AdS}\phi_{s'}^{\rm T})^{a_1\cdots a_{s'}}+\eta^{a_1a_2}(\delta_{AdS}\phi_{s'-2}^{\rm TT})^{a_3\cdots a_{s'}})\,\,=\cr\cr
&&\,\,\qquad\qquad\alpha_{a_1}\cdots\alpha_{a_{s'}}\Big[s'\partial^{a_1}\xi_{s'-1}^{a_2\cdots a_{s'}}\,\,+\,\,\sqrt{s^2-s^{\prime 2}\over 2s'+1}\left(\partial_Z+{s'\over Z}\right)\xi_{s'}^{a_1\cdots a_{s'}}\cr\cr
&&\,\,\qquad\qquad\qquad+\,\,{s'(s'-1)\over 2s'-3}\sqrt{s^2-(s'-1)^2\over 2s'-1}\left(\partial_Z-{s'-1\over Z}\right)\eta^{a_1a_2}\xi_{s'-2}^{a_3\cdots a_{s'}}\Big]\label{xispeqn}
\eea
for $s'\ge 2$,
\bea
\alpha_{a_1}\delta_{AdS}(\phi_{1}^{\rm T})^{a_1}\,\,=\,\,\alpha_{a_1}\Big[\partial^{a_1}\xi_0\,\,+\,\,\sqrt{s^2-1\over 3}\left(\partial_Z+{1\over Z}\right)\xi_{1}^{a_1}\Big]\label{xi1eqn}
\eea
for $s'=1$ and
\bea
\delta_{AdS}(\phi_{0}^{\rm T})\,\,=s\,\partial_Z\,\xi_{0}\label{xi0eqn}
\eea
for $s'=0$. From (\ref{xi0eqn}) we can use $\xi_0$ to set $\phi_0^{\rm T}=0$ and then from (\ref{xi1eqn}) we can use $\xi_1^a$ to set $(\phi_1^{\rm T})^a=0$. Next, using (\ref{xispeqn}) we can use $(\xi_{s'})^{a_1\cdots a_{s'}}$ to set $(\phi_{s'}^{\rm T})^{a_1\cdots a_{s'}}=0$ for $s'=2$, then $s'=3$ and so on up to $s'=s-1$.  Thus, in the end we use the complete set of gauge parameters to set
\bea
(\phi^T_{s'})^{a_1\cdots a_{s'}}&=&0\qquad\qquad s'\,\,=\,\,0,1,2,\cdots,s-1
\eea
These further gauge conditions can be written as
\bea
\bar{\alpha}^Z\,\Pi\, |\phi\rangle &=& 0
\eea
so that the operator defining the modified de Donder gauge condition (\ref{mdD}) becomes
\bea 
\bar{C}_{AdS}&\equiv&\bar{\alpha}\cdot\partial-\half\alpha\cdot\partial\bar\alpha^2-\half \alpha^Z \widetilde{e}_1\Bigl(\partial_Z+\frac{2s+d-5-2N_Z}{2Z}\Bigr) \bar\alpha\cdot\bar\alpha
\eea
Thus, in this gauge, the terms independent of $\alpha^a$ in (\ref{mdD}) imply that
\bea
\left(\partial_Z+{1\over Z}\right)\phi^{TT}_2(Z,x)&=&\left(\partial_Z+{1\over Z}\right)\, U_{3\over 2}(-1)^{N_Z}\phi^{TT}_2(x)\,\,=\,\, U_{1\over 2}(-1)^{N_Z}\phi^{TT}_2(x)\,\,=\,\,0\cr\cr
&&\quad\Rightarrow\phi^{TT}_2(x)\,\,=\,\,0
\eea
Making use of this result, the linear in $\alpha^a$ terms in (\ref{mdD}) imply that
\bea
\left(\partial_Z+{2\over Z}\right)(\phi^{TT}_3)^a(Z,x)&=&\left(\partial_Z+{2\over Z}\right)\, U_{5\over 2}(-1)^{N_Z}(\phi^{TT}_3)^a(x)
\,\,=\,\, U_{3\over 2}(-1)^{N_Z}(\phi^{TT}_3)^a(x)\,\,=\,\,0\cr\cr
&&\quad\Rightarrow(\phi^{TT}_3)^a(x)\,\,=\,\,0
\eea
Assume that $(\phi^{TT}_{s'})^{a_1\cdots a_{s'}}(x,Z)=0$. The terms of degree $s'-1$ in $\alpha^a$ in (\ref{mdD}) then imply that
\bea
\left(\partial_Z+{s'\over Z}\right)(\phi^{TT}_{s'+1})^{a_1\cdots a_{s'-1}}(Z,x)&=& \left(\partial_Z+{s'\over z}\right)\, U_{s'+\half}(-1)^{N_Z}(\phi^{TT}_{s'+1})^{a_1\cdots a_{s'-1}}(x)\cr\cr
&=& U_{s'-\half}(-1)^{N_Z}(\phi^{TT}_{s'+1})^{a_1\cdots a_{s'-1}}(x)\,\,=\,\,0\cr\cr
&&\quad\Rightarrow(\phi^{TT}_{s'+1})^{a_1\cdots a_{s'-1}}(x)\,\,=\,\,0
\eea
for $s'=3,4,\cdots,s-1$. Consequently, the only field that remains is $\phi_{s'}^{a_1\cdots a_s}$ and the modified de Donder gauge condition (\ref{mdD}) becomes the statement that this field is conserved
\bea
\partial_{a}\phi_{s}^{aa_2\dots a_s}=0\label{ConsHSCurr}
\eea
so that we recover the usual degrees of freedom in the spin $s$ primary current. It is straight forward to describe the action of the so(2,$d$) generators in this gauge. The representation is determined by the choice of $\Delta$ and $R^a$ given in (\ref{AdSRep}). Using the fact that in this gauge our state 
\bea
|\phi^{(s)}\rangle &=& {\alpha^{a_1}\cdots \alpha^{a_s}\over s!}\phi_{s}^{a_1\cdots a_s}|0\rangle\label{gaugefixedstate}
\eea
has no dependence on the $\alpha^Z$ oscillators, we easily find
\bea
\Delta |\phi^{(s)}\rangle &=& (s+d-2-N_Z){\alpha^{a_1}\cdots \alpha^{a_s}\over s!}\phi_{s}^{a_1\cdots a_s}|0\rangle= (s+d-2)|\phi^{(s)} \rangle
\eea
and
\bea
R^a_{(0)}|\phi^{(s)}\rangle &=&-Z\alpha^Z \bar{\alpha}^a|\phi^{(s)}\rangle\qquad
R^a_{(1)}|\phi^{(s)}\rangle\,\,=\,\,-{1\over 2}Z^2\partial^a|\phi^{(s)}\rangle
\eea
The contribution coming from the compensating gauge transformation is
\bea
R_{\rm comp}|\phi^{(s)}(x,Z)\rangle&=&\Bigg(\alpha\cdot\partial+\alpha^Z\Bigl(\partial_Z+\frac{2s+d-5}{2Z}\Bigr)\Bigg)ZU_{\nu+1}\bar{\alpha}^a(-1)^{N_Z}|\phi^{(s)}(x)\rangle\cr\cr\cr
&=&\alpha\cdot\partial ZU_{\nu+1}\bar{\alpha}^a(-1)^{N_Z}|\phi^{(s)}(x)\rangle+Z\alpha^Z\bar{\alpha}^a|\phi^{(s)}(x,Z)\rangle
\eea
The contribution with coefficient $\alpha^Z$ in the last term above cancels against the contribution from $R^a_{(0)}$. Thus, putting these contributions together, we find
\bea
R^a|\phi^{(s)}(x,Z)\rangle &=&-{1\over 2}Z^2\partial^a|\phi^{(s)}(x,Z)\rangle
+\alpha\cdot\partial \, Z\, U_{\nu+1}\, \bar{\alpha}^a\, (-1)^{N_Z}\, |\phi^{(s)}(x)\rangle
\eea
Since the action of $R^a$ does not introduce any dependence on $\alpha^Z$, the representation closes on the $\phi_{s}^{a_1\cdots a_s}$ fields, mirroring the analysis in the CFT.

When acting on the state (\ref{gaugefixedstate}) the equation of motion (\ref{HSeom}) becomes
\bea
\Bigl(\Box+\partial_Z^2-\frac{1}{Z^2} \left(\left(s-{1\over 2}\right)^2-\frac{1}{4}\right)\Bigr)|\phi^{(s)}(x,Z)\rangle=0
\eea

To derive the holographic mapping, it is convenient to assemble the complete collection of spinning fields into a single field\footnote{The states $|\phi^{(s)}\rangle$ all have dimension 1 independent of the spin $s$ so we can sensibly add them. This is easily seen by looking at (\ref{boundcond}) and noting that $Z$ is a length and we identify $|\phi^{(s)}(x)\rangle$ with a free CFT state of dimensions $s+1$. Further, at each $s$ there are only two independent and physical states that we index using $\pm$.}. Towards this end, introduce the vector
\bea
|\tilde{\phi}(x,Z)\rangle &=&\sum_{s=0}^\infty \left(e^{i(2s-{1\over 2})\varphi}|\phi_-^{(2s)}(x,Z)\rangle+e^{-i(2s-{1\over 2})\varphi}|\phi_+^{(2s)}(x,Z)\rangle\right)
\eea
The equation of motion can be written as
\bea
\Bigl(\Box+{\partial^2\over \partial Z^2}+\frac{1}{Z^2} \left({\partial^2\over\partial\varphi^2}+\frac{1}{4}\right)\Bigr)|\tilde\phi(x,Z)\rangle=0
\eea
To obtain an equation that naturally connects to the CFT we need to perform a small manipulation. Introduce a new state $|\tilde\phi\rangle=\sqrt{Z}|\phi\rangle$. The equation of motion for $|\phi\rangle$ is given by
\bea
\Bigl(\Box+{\partial^2\over \partial Z^2}+{1\over Z}\partial_Z+\frac{1}{Z^2}{\partial^2\over\partial\varphi^2}\Bigr)|\phi(x,Z)\rangle=0\label{nicebulkem}
\eea
We will see that the equation of motion in this form is directly connected to the equation of motion of the bilocal field $\eta$.

\subsection{Mapping}\label{covariantmapping}

\begin{figure}
\begin{center}
\tdplotsetmaincoords{75}{90}
\begin{tikzpicture}[tdplot_main_coords,line cap=round,line join=round,
    declare function={R=4;
    tcrit(\x)=180-atan2(sin(\tdplotmaintheta),cos(\tdplotmaintheta)*cos(\x));}]
 \foreach \Angle in {-75,-45,...,75}    
  {\pgfmathsetmacro{\tcrit}{tcrit(\Angle)}
  \draw[dashed] plot[variable=\t,domain=\tcrit:180,smooth] 
   ({R*cos(\Angle)*cos(\t)},{R*sin(\Angle)*cos(\t)},{R*sin(\t)}) ;}
 \draw[dashed] (0,-R) arc[start angle=-90,end angle=-270,radius=R];
 \begin{scope} 
  \clip (0,-R) arc[start angle=-90,end angle=90,radius=R]
   [tdplot_screen_coords] --(R,0) arc[start angle=0,end angle=180,radius=R];
  \shade[ball color=blue,fill opacity=0.6,tdplot_screen_coords] 
    circle[radius=R];
 \end{scope}
 \foreach \Angle in {-75,-45,...,75}    
  {\pgfmathsetmacro{\tcrit}{tcrit(\Angle)}
  \draw plot[variable=\t,domain=0:\tcrit,smooth] 
   ({R*cos(\Angle)*cos(\t)},{R*sin(\Angle)*cos(\t)},{R*sin(\t)}) ;}
 \draw (0,-R) arc[start angle=-90,end angle=90,radius=R]
  [tdplot_screen_coords] -- (R,0) arc[start angle=0,end angle=180,radius=R]
  -- cycle; 
  \filldraw[red] ({R*cos(45)*cos(0)},{R*sin(45)*cos(0)}) circle (0.2);
{\pgfmathsetmacro{\tcrit}{tcrit(45)}
  \draw[red, very thick] plot[variable=\t,domain=0:\tcrit,smooth] 
   ({R*cos(45)*cos(\t)},{R*sin(45)*cos(\t)},{R*sin(\t)}) ;}
  \filldraw[red] ({R*cos(-135)*cos(0)},{R*sin(-135)*cos(0)}) circle (0.2);
  {\pgfmathsetmacro{\tcrit}{tcrit(-135)}
  \draw[red, very thick] plot[variable=\t,domain=0:\tcrit,smooth] 
   ({R*cos(-135)*cos(\t)},{R*sin(-135)*cos(\t)},{R*sin(\t)}) ;}
\end{tikzpicture}   
\caption{The base of the hemisphere above is a disk centred at $\left({x_1+x_2\over 2},{y_1+y_2\over 2}\right)$ on the horizontal plane, parametrized by $X,Y$, which is the boundary on which the CFT is defined. The vertical direction, perpendicular to the boundary, is parametrized by the emergent holographic coordinate $Z$. The lines of longitude shown correspond to lines of fixed $\varphi$. Adjusting the energies $p_1^0$ and $p_2^0$ of the two particles in the bilocal moves the bulk excitation dual to the bilocal along the lines of longitude. To move between the lines of longitude we need to change $\varphi$ which rotates the $(Z_1,Z_2)$ coordinates. The two red circles on the plane show the location of the excitations in the bilocal. The excitation is located at any point on the red curve in the bulk. Choosing a specific $p_1^0$ and $p_2^0$ will localize the excitation on this curve.}\label{RTsurface}
\end{center}
\end{figure}
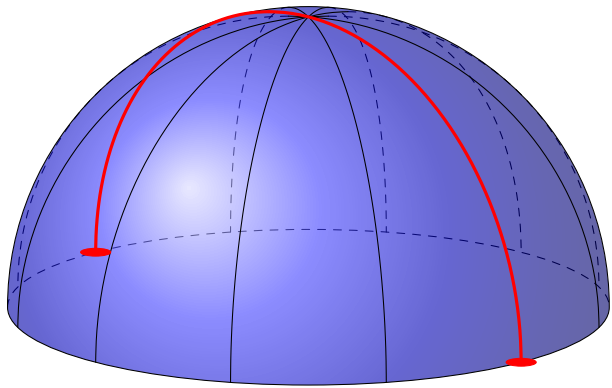

The discussion of the previous two sections has been valid for general $d$. We now specialize to $d=3$. Our strategy is to use the insights obtained from the light front map. In that case the gravity coordinates $X$ and $X^-$ take a centre of mass form, with $p_i^+$ play the role of the mass for the field at $x_i$. It is natural to expect a similar formula for $X$ and $Y$ with the role of the mass now naturally played by the energies $p_1^0$ and $p_2^0$. This motivates the formulas
\bea
X={p_1^0 x_1+p_2^0 x_2\over p_1^0+p_2^0} \qquad\qquad
Y={p_1^0 y_1+p_2^0 y_2\over p_1^0+p_2^0}\label{commap2grav1}
\eea
When writing these formulas we have in mind bilocal fields composed of excitations that are described by wave packets tightly peaked about the energies $p_1^0$ and $p_2^0$. The $Z$-coordinate is determined by requiring that we obtain the correct entanglement wedge for a subregion given by a disk in the $X,Y$ plane at $Z=0$. The RT surface for a disk subregion in AdS$_4$ is a hemisphere. To obtain the correct entanglement wedge, the map must satisfy (refer to Figure \ref{RTsurface})
\bea
\left(X-{x_1+x_2\over 2}\right)^2+\left(Y-{y_1+y_2\over 2}\right)^2+Z^2={(x_1-x_2)^2+(y_1-y_2)^2\over 4}
\eea
This is easily solved by setting $Z^2=Z_1^2+Z_2^2$ where
\bea
Z_1={\sqrt{p_1^0p_2^0}\over p_1^0+p_2^0}(x_1-x_2)\qquad\qquad
Z_2={\sqrt{p_1^0p_2^0}\over p_1^0+p_2^0}(y_1-y_2)\label{commap2grav2}
\eea
The angle $\theta$ in the light front map, which plays the role of a polarization, is a local angle in the $X,Z$ plane, defined with respect to an origin defined by the bilocal and not obviously related to any angle defined in the CFT. This is simply a consequence of the fact that choosing light cone gauge and then solving the constraint does not preserve the Poincar\'{e} subgroup of so(2,3), so the relation between rotations and boosts in the CFT and the dual gravity is not straight forward. Here we introduce an angle $\varphi$ that will play a similar role. Since our description of the higher spin gravity and the CFT are both covariant with respect to the boundary Poincar\'{e} symmetry, we expect a simple relation between $\varphi$ and angles in the CFT. With this in mind, it is natural to identify
\bea
Z_1=Z\cos\varphi\qquad Z_2=Z\sin\varphi
\eea
Notice that once again $Z$ only vanishes when the two fields in the bilocal are coincident. This implies that the single trace primaries are again localized to a neighbourhood of the boundary and further, that bilocals composed using two well separated fields, are located deep in the bulk. By making use of the OPE operators localized deep in the bulk can be expressed as elements of the boundary algebra which explains how our map is consistent with the principle of the holography of information. Thus, with the above map the resulting collective field theory localizes information exactly as expected in a theory of quantum gravity.

The above map is invertible with the result
\bea
x_1=X+\sqrt{p_2^0\over p_1^0}Z_1\qquad x_2=X-\sqrt{p_1^0\over p_2^0}Z_1\label{commap2cft1}
\eea
\bea
y_1=Y+\sqrt{p_2^0\over p_1^0}Z_2\qquad y_2=Y-\sqrt{p_1^0\over p_2^0}Z_2
\label{commap2cft2}
\eea
Expressing the momenta in the form $p_x=-i{\partial\over \partial x}$ and using (\ref{commap2grav1}) and (\ref{commap2grav2}), as well as (\ref{commap2cft1}) and (\ref{commap2cft2}), the chain rule can be used to derive the mapping between bulk and CFT momenta. The result is
\bea
P^X&=&p^x_1+p^x_2\qquad\qquad\qquad\qquad\quad P^Y\,\,=\,\,p^y_1+p^y_2\cr\cr
P_{Z_1}&=&\sqrt{p_2^0\over p_1^0}p_1^x-\sqrt{p_1^0\over p_2^0}p_2^x\qquad\qquad P_{Z_2}\,\,=\,\,\sqrt{p_2^0\over p_1^0}p_1^y-\sqrt{p_1^0\over p_2^0}p_2^y\label{mommap2cft}
\eea
and
\bea
p_1^x&=&{p^0_1\over p_1^0+p_2^0}P^X+{\sqrt{p_1^0p_2^0}\over p_1^0+p_2^0}P_{Z_1}\qquad\qquad p_1^y\,\,=\,\,{p^0_1\over p_1^0+p_2^0}P^Y+{\sqrt{p_1^0p_2^0}\over p_1^0+p_2^0}P_{Z_2}\cr\cr
p_2^x&=&{p^0_2\over p_1^0+p_2^0}P^X-{\sqrt{p_1^0p_2^0}\over p_1^0+p_2^0}P_{Z_1}\qquad\qquad p_2^y\,\,=\,\,{p^0_2\over p_1^0+p_2^0}P^Y-{\sqrt{p_1^0p_2^0}\over p_1^0+p_2^0}P_{Z_2}
\eea
The formulas for $P^X$ and $P^Y$ given above are exactly what they should be: the boundary CFT and bulk gravity share translation invariance in both $X$ and $Y$. These formulas simply equate the conserved charges of these symmetries. Below we will argue that the formulas for $P_{Z_i}$ lead to the correct bulk equations of motion. The last ingredient in the map is the identification of the bilocal field $\eta$ and the gravity field $|\phi\rangle$. We consider the equal time bilocal theory, described by the field $\eta(t,\vec{x}_1,\vec{x}_2)$. In the dual gravity, we reduce to physical degrees of freedom by using (\ref{ConsHSCurr}) to eliminate the temporal polarizations of the current. This gives a field $|\phi(X^A)\rangle\equiv|\phi(t,X,Y,\alpha^1,\alpha^2)\rangle$ where we have explicitly indicated the dependence on the oscillators. The map between fields is
\bea
\eta(t,X+\sqrt{p_2^0\over p_1^0}Z_1,Y+\sqrt{p_2^0\over p_1^0}Z_2,X-\sqrt{p_1^0\over p_2^0}Z_1,Y-\sqrt{p_1^0\over p_2^0}Z_2) =|\phi(t,X,Y,Z_1,Z_2)\rangle \label{fieldident}
\eea
Does this map is provide a valid bulk reconstruction? To verify that the bulk fields obey the correct equations of motion we argue that the CFT equations of motion imply the bulk equations of motion. The CFT equations of motion, after Fourier transforming to momentum space and using (\ref{fieldident}), are given by
\bea
((p_1^0)^2-(p_1^x)^2-(p_1^y)^2)|\phi(P^A)\rangle =0=((p_2^0)^2-(p_2^x)^2-(p_2^y)^2)|\phi(P^A)\rangle\label{CFTeom}
\eea
while the bulk equation of motion is given by (\ref{nicebulkem}). Rewriting in terms of $Z_1$ and $Z_2$, the equation of motion for $|\phi(X^A)\rangle$ is given by
\bea
\Bigl(-{\partial^2\over\partial t^2}+\partial_X^2+\partial_Y^2+\partial_{Z_1}^2+\partial_{Z_2}^2\Bigr)|\phi(X^A)\rangle=0
\eea
After a Fourier transform to momentum space, we have
\bea
\Bigl((P^0)^2-(P^X)^2-(P^Y)^2-P_{Z_1}^2-P_{Z_2}^2\Bigr)|\phi(P^A)\rangle =0\label{eqntoprove}
\eea
Finally, since the boundary CFT and the bulk gravity share the same time translation invariance we know that $P^0=p_1^0+p_2^0$. Start with the LHS of (\ref{eqntoprove}) and use (\ref{mommap2cft}) to find
\bea
\Bigl((P^0)^2-{p_1^0+p_2^0\over p_1^0}((p^x_1)^2+(p^y_1)^2)-{p_1^0+p_2^0\over p_2^0}((p^x_2)^2+(p^y_2)^2)\Bigr)|\phi\rangle
\label{BEOM}
\eea
Enforcing the CFT equations of motion (\ref{CFTeom}) we find
\bea
\Bigl((P^0)^2-{p_1^0+p_2^0\over p_1^0}(p_1^0)^2-{p_1^0+p_2^0\over p_2^0}(p_2^0)^2\Bigr)|\phi\rangle=\Big((P^0)^2-(p_1^0+p_2^0)^2\Big)|\phi\rangle
\eea
which does indeed vanish. This completes the demonstration that the CFT equations of motion, together with the holographic mapping, imply the bulk equations of motion.

To complete this discussion consider the boundary conditions obeyed by the field, which are spelled out in (\ref{boundcond}). After reducing to spatial polarizations, only two independent components of the current (denoted $j^{\pm}_{(2s)}$) at each spin remain. The OPE (\ref{explicitOPE}) becomes
\bea
\eta(t,\vec{x}_1,\vec{x}_2)&=&
\sum_{s=0}^\infty\sum_{d=0}^\infty c_{sd} 
\left(y^i {\partial\over\partial x^i}\right)^{2d}\left(
(y^1+iy^2)^{2s}\,\, j_{(2s)}^{+}(t,\vec{x})+(y^1-iy^2)^{2s}\,\, j_{(2s)}^{-}(t,\vec{x})\right)
\nonumber
\eea 
Using the holographic mapping (\ref{commap2cft1}) and (\ref{commap2cft2}) we easily find
\bea
y^1=x_1-x_2={p_2^0+p_1^0\over\sqrt{p_1^0 p_2^0}}Z_1=
{p_2^0+p_1^0\over\sqrt{p_1^0 p_2^0}}Z\cos\varphi
\eea
\bea
y^2=y_1-y_2={p_2^0+p_1^0\over\sqrt{p_1^0 p_2^0}}Z_2=
{p_2^0+p_1^0\over\sqrt{p_1^0 p_2^0}}Z\sin\varphi
\eea
\bea
 x^1=X+O(Z)\qquad y^1=Y+O(Z)
\eea
so that, as $Z\to 0$ we have
\bea
\eta(t,\vec{x}_1,\vec{x}_2)=
\sum_{s=0}^\infty c_{s0} Z^{2s}{(p_2^0+p_1^0)^{2s}\over (p_1^0 p_2^0)^s}\left(
e^{2is\varphi}\,\, j_{(2s)}^{+}(t,X,Y)+e^{-2is\varphi}\,\, j_{(2s)}^{-}(t,X,Y)+O(Z)\right)
\nonumber
\eea 
in perfect agreement with (\ref{boundcond}). This proves that we have indeed obtain a valid bulk reconstruction.

Finally, it is interesting to study the behaviour of the extra holographic coordinate $Z$ of a bulk excitation dual to a given bilocal. The coordinate $Z$ is given by
\bea
Z={\sqrt{p_1^0p_2^0}\over p_1^0+p_2^0}\sqrt{(x_1-x_2)^2+(y_1-y_2)^2}
\eea
Since energies are always positive we have $0<{\sqrt{p_1^0p_2^0}\over p_1^0+p_2^0}<1$. This factor is maximized when $p_1^0=p_2^0$, so that the bulk excitation is located deepest in the AdS spacetime when the total energy is shared equally between the two excitations in the bilocal. If either of the excitations carry most of the energy this factor becomes small and the excitation is located close to the boundary. Finally, to locate excitations deep in the bulk, we need a large spacial separation $\sqrt{(x_1-x_2)^2+(y_1-y_2)^2}$ between the two fields in the bilocal. 

\section{Conclusions}\label{conclusions}

In this article we have used collective field theory \cite{Jevicki:1979mb,Jevicki:1980zg} as a constructive approach to the holography of vector models \cite{Das:2003vw}. First we performed a change from the original scalar field to gauge invariant bilocal field variables. Then we performed a change of spacetime coordinates and specified the relation between the fluctuation of the bilocal field and the dual higher spin gravity fields. This provides a mapping between the complete set of single trace primaries and the complete set of bulk scalar plus spinning gauge fields. There are a number of lessons about bilocal holography that are worth listing:
\begin{itemize}
\item[1.] Bilocal holography provides a complete reconstruction of the bulk spacetime. The equation of motion for every bulk field, together with the correct boundary condition for each of these fields, is reproduced by the bilocal holography map.
\item[2.] The choice of a gauge in the higher spin gravity is related to the choice of bilocal used in the CFT description. Indeed, choosing an equal time bilocal in the CFT is a choice about how the CFT will be reduced to independent degrees of freedom: the temporal polarization will be eliminated. The choice of gauge in gravity is also related to choosing how the theory will be reduced: the gauge condition and its constraint both eliminate degrees of freedom. In this way, the equal $x^+$ bilcoal is naturally related to lightcone gauge, while equal time bilocal is naturally related to temporal gauge.
\item[3.] Both in the lightfront and the covariant version of bilocal holography, the choice of the momenta associated to the holographic coordinate $Z$ is fixed by the bulk equations of motion. The bulk equations of motion determine how the fields in the bulk $Z>0$ are determined in terms of their boundary values. The boundary condition (which is set at $Z=0$) encodes the usual GKPW map of AdS/CFT and it determines the structure of a local angle which plays the role of a polarization, packaging the spinning fields into a single field.
\item[4.] The holographic map constructs the extra holographic radial coordinate $Z$ as the distance between the two operators in the bilocal. Single trace primaries which are obtained by taking derivatives and letting the two fields in the bilocal approach a common point, live in a neighbourhood of the boundary at $Z=0$. Bilocals with well separated fields are located deep in the bulk at some $Z\gg 0$. The OPE expresses bilocals (and their products) in terms of single trace primaries (and their products), so that our formula for $Z$ is perfectly consistent with the principle of the holography of information. Collective field theory provides a geometrization of the space of CFT operators in a manner that is in perfect agreement with how we expect information to localize in a theory of quantum gravity. 
\item[5.] For the case of a strip subregion (for the equal $x^+$ bilocal) or of a disk subregion (for the equal $t$ bilocal) we find that the map of bilocal holography predicts the correct entanglement wedge. This is further evidence that collective field theory localizes information exactly as in a theory of quantum gravity.
\end{itemize}

Point 4 deserves some extra discussion. Here we are constructing a 4d gravity theory from a 3d CFT. The CFT does not have enough degrees of freedom to produce a genuine 4d theory, so there must be redundancies between the degrees of freedom of the 4d theory. Our analysis shows that the collective field theory description does indeed have redundancies and they take exactly the form predicted by the holography of information. This is strong evidence that collective field theory is producing a higher dimensional theory of gravity.

There are a number of ways in which this work can be extended. First, it would be interesting to compute subleading corrections in ${1\over N}$ and make contact with the interaction vertices of higher spin gravity. These vertices are generated in the CFT by expanding the Jacobian about the leading large $N$ configuration $\sigma_0$.

Moving on to other backgrounds, constructing the holographic map for the equal time description of the CFT at finite temperature, would provide deep insights into the geometry dual to the thermofield double state. This is particularly interesting given the fact that in this case we expect horizons in the bulk spacetime. 

Finally, it is interesting to ask how this map for the vector model can be extended to a theory of free matrices. Again, we can declare that it is the singlet sector that is dual to the gravity theory. For the matrix model the space of invariants is much richer that it was for the vector model. For the vector model we could only produce bilocal fields because the only way to produce an O(N) is invariant is by contracting a pair of vectors. For the matrix model we can produce a U(N) invariant by taking a trace of any number $k$ of matrices and hence we generate $k$-local invariant collective fields for every $k$. Correctly constructing the holographic map in this setting would be another convincing test of the idea that collective field theory provides a constructive approach to holography.

\begin{center} 
{\bf Acknowledgements}
\end{center} 
This research is supported by a start up research fund of Huzhou University, a Zhejiang Province talent award and by a Changjiang Scholar award. We thank Antal Jevicki for very useful discussions on the subject of this paper.

\begin{appendix}

\section{Identities obeyed by the bilocal field}\label{collectiveidentity}

In Section \ref{ETbilocal} we made use of identities obeyed by the equal time (and equal $x^+$) collective bilocal field. In this Appendix we derive these identities. The identity obeyed by the equal time bilocal follows by evaluating
\bea
\eta^{\mu\nu}{\partial\over\partial y^\mu}{\partial\over\partial x^\nu}
:\phi^a(x^\mu+y^\mu)\phi^a(x^\mu-y^\mu):
\eea
If both derivatives act on a single field we have
\bea
\eta^{\mu\nu}{\partial\over\partial y^\mu}{\partial\over\partial x^\nu}
\phi^a(x^\mu\pm y^\mu)=\pm \eta^{\mu\nu}{\partial\over\partial x^\mu}{\partial\over\partial x^\nu}\phi^a(x^\mu\pm y^\mu)=0
\eea
where we use the free equation of motion. Thus
\bea
&&\eta^{\mu\nu}{\partial\over\partial y^\mu}{\partial\over\partial x^\nu}
:\phi^a(x^\mu+y^\mu)\phi^a(x^\mu-y^\mu):\cr\cr
&&\qquad =\,\,
\eta^{\mu\nu}:{\partial \phi^a(x^\mu+y^\mu)\over\partial y^\mu}
{\partial \phi^a(x^\mu-y^\mu)\over\partial x^\nu}:+
\eta^{\mu\nu}:{\partial \phi^a(x^\mu+y^\mu)\over\partial x^\nu}
{\partial\phi^a(x^\mu-y^\mu)\over\partial y^\mu}:\cr\cr
&&\qquad =\,\,\eta^{\mu\nu}:{\partial \phi^a(x^\mu+y^\mu)\over\partial x^\mu}
{\partial \phi^a(x^\mu-y^\mu)\over\partial x^\nu}:-\eta^{\mu\nu}:{\partial \phi^a(x^\mu+y^\mu)\over\partial x^\nu}{\partial\phi^a(x^\mu-y^\mu)\over\partial x^\mu}:\cr\cr
&&\qquad =\,\, 0\label{KeyIdent}
\eea
Now, work in Cartesian coordinates, specialize to $d=3$ and evaluate the identity (\ref{KeyIdent}) at $y^0=0$
\bea
&&{\partial\over\partial t}\Big(:\partial_t\phi^a(t,\vec{x}+\vec{y})\phi^a(t,\vec{x}-\vec{y}):-:\phi^a(t,\vec{x}+\vec{y})\partial_t\phi^a(t,\vec{x}-\vec{y}):\Big)\cr\cr
&=&({\partial\over\partial y^1}{\partial\over\partial x^1}+{\partial\over\partial y^2}{\partial\over\partial x^2}):\phi^a(x^+,\vec{x}+\vec{y})\phi^a(x^+,\vec{x}-\vec{y}):
\eea
which implies that
\bea
&&\Big(:\partial_t\phi^a(t,\vec{x}+\vec{y})\phi^a(t,\vec{x}-\vec{y})
-\phi^a(t,\vec{x}+\vec{y})\partial_t\phi^a(t,\vec{x}-\vec{y}):\Big)\cr\cr
&&={\partial_{y^1}\partial_{x^1}+\partial_{y^2}\partial_{x^2}\over\partial_t}
:\phi^a(t,\vec{x}+\vec{y})\phi^a(t,\vec{x}-\vec{y}):\label{equaltimeidentity}
\eea
This is the first identity we wanted to prove.

For the identity obeyed by the equal $x^+$ bilocal field, work in lightcone coordinates and evaluate the identity (\ref{KeyIdent}) at $y^+=0$
\bea
&&-{\partial\over\partial x^-}
\Big(:\partial_+\phi^a(x^+,x^-+y^-,x+y)\phi^a(x^+,x^--y^-,x-y)\cr\cr
&&\qquad\qquad\qquad
-\phi^a(x^+,x^-+y^-,x+y)\partial_+\phi^a(x^+,x^--y^-,x-y):
\Big)\cr\cr
&&=({\partial\over\partial y^-}{\partial\over\partial x^+}+
{\partial\over\partial y}{\partial\over\partial x})
:\phi^a(x^+,x^-+y^-,x+y)\phi^a(x^+,x^--y^-,x-y):
\eea
which implies that
\bea
&&-
\Big(:\partial_+\phi^a(x^+,x^-+y^-,x+y)\phi^a(x^+,x^--y^-,x-y)\cr\cr
&&\qquad\qquad\qquad
-\phi^a(x^+,x^-+y^-,x+y)\partial_+\phi^a(x^+,x^--y^-,x-y):
\Big)\cr\cr
&&={\partial_{y^+}\partial_{x^-}+\partial_y\partial_x\over \partial^+}
:\phi^a(x^+,x^-+y^-,x+y)\phi^a(x^+,x^--y^-,x-y):\label{xplusidentity}
\eea
This is the second identity we wanted to prove.

\section{Further comments on higher spin gravity}

In our discussion of higher spin gravity in the light cone gauge, following \cite{Metsaev:1999ui} we have used the double-traceless Fronsdal field $\Phi^{A_1\ldots A_s}$
\cite{Fronsdal:1978vb}. The description used in Section \ref{covariant}, developed in \cite{Metsaev:2008fs}, uses double-traceless so($d$-1,1) algebra fields. This is not the Fronsdal field and is used because it facilitates the connection to the dual CFT. In this Appendix we will simply state the relation between these two descriptions and refer the reader to \cite{Metsaev:2008ks} for the details. Denoting the Fock space state corresponding the spin $s$ Fronsdal field by $\Phi$ and the Fock space state for the spin $s$ field of the so($d$-1,1) description by $\phi$, the relation is

\bea \phi  =  Z^{\frac{1-d}{2}}{\cal N}  \Pi^{\phi\Phi}\Phi \eea
where
\bea
\Pi^{\phi\Phi} &\equiv& \Pi_\alpha^{[1]} + \alpha^2 \frac{1}{2(2N_\alpha + d)}
\Pi_\alpha^{[1]} (\bar\alpha^2 + \frac{2N_\alpha +d}{2N_\alpha + d-2}\bar\alpha^Z\bar\alpha^Z)
\eea
\bea
\Pi_\alpha^{[1]} \equiv \Pi^{[1]}(\alpha,0,N_\alpha,\bar\alpha,0,d)
\eea
\bea 
{\cal N}\equiv\Bigl(\frac{2^{N_Z}\Gamma(N_\alpha+N_Z+\frac{d-3}{2})\Gamma( 2N_\alpha+d-3)}{\Gamma(N_\alpha+\frac{d-3}{2})\Gamma(2N_\alpha+N_Z+d-3)}\Bigr)^{1/2}
\eea
and
\bea
\Pi^{[1]}(\alpha,\alpha^Z, X,\bar\alpha,\bar\alpha^Z, Y) \equiv \sum_{n=0}^\infty (\alpha^2+\alpha^Z\alpha^Z)^n \frac{(-)^n \Gamma(X+\frac{Y-2}{2} + n)}{4^n n! \Gamma(X + \frac{Y-2}{2} +2n)}(\bar\alpha^2+\bar\alpha^z\bar\alpha^z)^n
\eea
The inverse transformation is
\bea 
\Phi&=&Z^{\frac{d-1}{2}} \Pi^{\Phi\phi} {\cal N} \phi
\eea
where
\bea
\Pi^{\Phi\phi}&\equiv& \Pi_{\bf\alpha}^{[1]}+ \alpha^A\alpha_A \frac{1}{2(2 (N_\alpha+N_Z+d+1)}\Pi_{\bf\alpha}^{[1]} \left(\alpha^A\alpha_A-\frac{2}{2 (N_\alpha+N_Z) + d-1}\bar\alpha^Z\bar\alpha^Z\right)\cr\cr\cr
\Pi_{\bf\alpha}^{[1]} &\equiv&\Pi^{[1]}(\alpha,\alpha^Z,N_\alpha+N_Z,\bar\alpha,\bar\alpha^z,d+1)
\eea
In this Appendix $\alpha^2\equiv\alpha^a\alpha^b\eta_{ab}$ and $\bar\alpha^2\equiv\bar\alpha^a\bar\alpha^b\eta_{ab}$.

\end{appendix}

\end{document}